\documentclass[utf8]{frontiersSCNS} 

\usepackage{url,hyperref,lineno,microtype,subcaption}
\usepackage[onehalfspacing]{setspace}

\usepackage{graphicx}

\usepackage{epstopdf}

\usepackage{xr-hyper}

\usepackage{bbold}

\def\keyFont{\fontsize{8}{11}\helveticabold }

\def\firstAuthorLast{Fabian Schubert and Claudius Gros} 

\def\Authors{Fabian Schubert\,$^{1,*}$ and Claudius Gros\,$^{1}$}

\def\Address{$^{1}$Institute for Theoretical Physics, 
Goethe University Frankfurt am Main, Germany}

\def\corrAuthor{Institute for Theoretical Physics\\
  Goethe University Frankfurt am Main\\
  Max-von-Laue-Str. 1\\
  60438 Frankfurt am Main, Germany}

\def\corrEmail{fschubert@itp.uni-frankfurt.de}

\begin{document}

\onecolumn
\firstpage{1}

\title[Homeostatic regulation of echo-state networks]
{Local homeostatic regulation of the spectral radius of echo-state networks}

\author[\firstAuthorLast ]{\Authors} 
\address{} 
\correspondance{} 

\extraAuth{}%

\maketitle

\begin{abstract}
\section{}
Recurrent cortical networks provide reservoirs of states that are thought to play a crucial role for sequential information processing in the brain.  However, classical reservoir computing requires manual adjustments of global network parameters, particularly of the spectral radius of the recurrent synaptic weight matrix. It is hence not clear if the spectral radius is accessible to biological neural networks.

Using random matrix theory, we show that the spectral radius is related to local properties of the neuronal dynamics whenever the overall dynamical state is only weakly correlated. This result allows us to introduce two local homeostatic synaptic scaling mechanisms, termed flow control and variance control, that implicitly drive the spectral radius towards the desired value. For both mechanisms the spectral radius is autonomously adapted while the network receives and processes inputs under working conditions.

We demonstrate the effectiveness of the two adaptation mechanisms under different external input protocols.  Moreover, we evaluated the network performance after adaptation by training the network to perform a time-delayed XOR operation on binary sequences.  As our main result, we found that flow control reliably regulates the spectral radius for different types of input statistics. Precise tuning is however negatively affected when interneural correlations are substantial. Furthermore, we found a consistent task performance over a wide range of input strengths/variances.  Variance control did however not yield the desired spectral radii with the same precision, being less consistent across different input strengths.

Given the effectiveness and remarkably simple mathematical form of flow control, we conclude that self-consistent local control of the spectral radius via an implicit adaptation scheme is an interesting and biological plausible alternative to conventional methods using set point homeostatic feedback controls of neural firing.

\tiny
\keyFont{ \section{Keywords:} recurrent networks, homeostasis,
    synaptic scaling, echo-state networks,
    reservoir computing, spectral radius} 
\end{abstract}

\bigskip

\section{Introduction}
\label{sect:introduction}

Cortical networks are highly recurrent, a property
that is considered to be crucial for processing and
storing temporal information. For recurrent networks
to remain stable and functioning, the neuronal
firing activity has to be kept within a certain
range by autonomously active homeostatic mechanisms.
It is hence important to study homeostatic mechanisms
on the level of single neurons, as well as the more
theoretic question of characterizing the dynamic
state that is to be attained on a global network
level. It is common to roughly divide adaptation mechanisms
into intrinsic homeostasis, synaptic
homeostasis, and metaplasticity.

Synaptic scaling was identified as a mechanism
that can postsynaptically regulate neural firing
by adjusting synaptic efficacies in a proportional,
multiplicative way. This finding has led to numerous
studies investigating the role of synaptic scaling
in controlling neural network activity
\citep{Turrigiano_1998,Turrigiano_2000,Turrigiano_2008}
and in stabilizing other plasticity mechanisms
\citep{vanRossum_2000,Stellwagen2006,Tetzlaff2011,Toyoizumi2014}.
Indeed, synaptic scaling has proven successful in
stabilizing activity in recurrent neural networks
\citep{Lazar_2009,Remme2012,Zenke2013,Effenberger_2015,Miner_2016}.
However, these studies either used synaptic scaling as
the sole homeostatic mechanism \citep{Zenke2013,Remme2012}
or resorted to a variant of synaptic scaling where
the scaling is not dynamically determined through a
control loop using a particular target activity,
but rather by a fixed multiplicative normalization rule
\citep{Lazar_2009,Effenberger_2015,Miner_2016}.
Therefore, these homeostatic models cannot account
for higher moments of temporal activity patterns, i.e.,
their variance, as this would require at least the
tuning of two parameters \citep{cannon2017stable}.

Within more abstract models of rate encoding neurons,
intrinsic homeostasis and synaptic scaling
essentially correspond to adjusting a bias and
gain factor on the input entering a nonlinear
transfer function. Within this framework,
multiple dual-homeostatic adaptation rules have been
investigated concerning their effect on network performance.
In this framework, the adaptation of the bias acts
as an intrinsic plasticity mechanism for the control
of the internal excitability of a neuron
\citep{Franklin_1992,Abbott_1993,Borde_1995},
while the gain factors functionally correspond to a
synaptic scaling of the recurrent weights.
Learning rules for these types of models were
usually derived by defining a target output
distribution that each neuron attempts to reproduce
by changing neural gains and biases
\citep{Triesch_2007,steil2007intrinsicplasticity,
schrauwen2008improving,boedecker2009initialization},
or were directly derived from an
information-theoretic measure \citep{Bell_1995}.

While these studies did indeed show performance
improvements by optimizing local information
transmission measures, apparently, optimal
performance can effectively be traced back to
a global parameter, the spectral radius of
the recurrent weight matrix \citep{schrauwen2008improving}.
Interestingly, to our knowledge, theoretical studies
on spiking neural networks did not explicitly
consider the spectral radius as a parameter
affecting network dynamics. Still, the theory of
balanced states in spiking recurrent networks
established the idea that synaptic strengths
should scale with $1/\sqrt{k}$, where $k$ is the
average number of afferent connections \citep{VanVreeswijk1998}.
According to the circular law of random matrix theory,
this scaling rule simply implies that the spectral
radius of the recurrent weight matrix remains finite
as the number of neurons $N$ increases.
More recent experiments on cortical cultures
confirm this scaling \citep{Barral2016}.

In the present study, we investigated whether the spectral
radius of the weight matrix in a random recurrent
network can be regulated by a combination of
intrinsic homeostasis and synaptic scaling. Following
the standard echo-state framework, we used rate encoding
tanh-neurons as the model of choice. However,
aside from their applications as efficient machine
learning algorithms, echo state networks are potentially
relevant as models of information processing in the brain
\citep{nikolic2009distributed,Hinaut_2015,enel2016reservoir}.
Note in this context that extensions to layered ESN
architectures have been presented by \citet{gallicchio2017echo},
which bears a somewhat greater resemblance to the hierarchical
structure of cortical networks than the usual shallow ESN
architecture. This line of research illustrates the importance
of examining whether local and biological plausible principles
exist that would allow to tune the properties of the neural
reservoir to the ``edge of chaos" \citep{livi2018determination},
particularly when a continuous stream of inputs is present.
The rule has to be independent of both the network topology,
which is not locally accessible information, and the
distribution of synaptic weights.

We propose and compare two unsupervised homeostatic mechanisms,
which we denote by flow control and variance control. Both
regulate the mean and variance of neuronal firing such that
the network works in an optimal regime concerning sequence
learning tasks. The mechanisms act on two sets of node-specific
parameters, the biases $b_i$, and the neural gain factors $a_i$.
We restricted ourselves to biologically plausible adaptation
mechanisms, viz adaptation rules for which the dynamics of
all variables are local, i.e., bound to a specific neuron.
Additional variables enter only when locally accessible.
In a strict sense, this implies that local dynamics are
determined exclusively by the neuron's dynamical variables
and by information about the activity of afferent neurons.
Being less restrictive, one could claim that it should also
be possible to access aggregate or mean-field quantities
that average a variable of interest over the population.
For example, nitric oxide is a diffusive neurotransmitter
that can act as a measure for the population average of
neural firing rates \citep{Sweeney_2015}.

Following a general description of the network model,
we introduce both adaptation rules and evaluate their
effectiveness in tuning the spectral radius in Section
\ref{sect:flow_control_results} and \ref{sect:variance_control_results}.
We assess the performance of networks that were subject
to adaptation in Section \ref{sect:XOR}, using a nonlinear
sequential memory task. Finally, we discuss the influence of
node-to-node cross-correlations within the population
in Section \ref{sect:correlations}.

\section{Results}

\bigskip\subsection{Model}
\label{sect_model}

A full description of the network model and parameters can be
found in the methods section. We briefly introduce the network
dynamics as
%
  \begin{align}
  x_i(t) &= x_{{\rm r},i}(t) + I_i(t) \label{x_i_introduction} \\
  x_{{\rm r},i}(t) &:= a_i\sum_{j=1}^N W_{ij} y_j(t-1)  \label{x_r_i_introduction}\\
  y_i(t) &= \tanh\left(x_i(t) - b_i\right) \; . \label{y_introduction}
  \end{align}
%
Each neuron's membrane potential $x_i$ consists of a recurrent
contribution $x_{{\rm r},i}(t)$ and an external
input $I_i(t)$. The biases $b_i$ are subject to the following
homeostatic adaptation:
%
  \begin{equation}
  b_i(t)= b_i(t-1) + \epsilon_{\rm b} \left[y_i(t) -
  \mu_{\rm t} \right] \; .
  \label{b_i_introduction}
  \end{equation}
%
Here, $\mu_{\rm t}$ defines a target for the average activity
and $\epsilon_{\rm b}$ is the adaptation rate.

The local parameters $a_i$ act as scaling factors on the recurrent weights.
We considered two different forms of update rules. Loosely speaking,
both drive the network towards a certain dynamical state which
corresponds to the desired spectral radius. The difference between
them lies in the variables characterizing this state:
While flow control defines a relation between the
variance of neural activity and the variance of the total
recurrent synaptic current, variance control does so by a more
complex relation between the variance of neural activity and
the variance of the synaptic current from the external input.

\bigskip

\subsubsection{Flow control}

The first adaptation rule, flow control, is given by
%
  \begin{equation}
  a_i(t) = a_i(t-1)\Big[1+ \epsilon_{\rm a} \Delta
  R_i(t)\Big],
  \quad\quad
  \Delta R_i(t) = R_{\rm t}^2 y_i^2(t-1) - x_{{\rm
      r},i}^2(t)\;.
  \label{a_i_flow_introduction}
  \end{equation}
%
The parameter $R_{\rm t}$ is the desired target spectral radius
and $\epsilon_{\rm a}$ the adaptation rate of the scaling factor.
The dynamical variables $y_i^2$ and $x_{{\rm r},i}^2$ have
been defined before in
Eqs.~(\ref{x_i_introduction}) and (\ref{x_r_i_introduction}).
We also considered an alternative global update rule
where $\Delta R_i(t)$ is given by
%
  \begin{equation}
  \Delta R_i(t) = \frac{1}{N}\Big[
  R_{\rm t}^2\,{||\mathbf{y}(t-1)||}^2-
  {||\mathbf{x}_{\rm r}(t)||}^2 \Big] \; ,
  \label{delta_R_global_introduction}
  \end{equation}
%
where $|| \cdot ||$ denotes the euclidean vector norm. However, since
this is a non-local rule, it only served as a comparative model to
Eq.~(\ref{a_i_flow_introduction}) when we investigated the
effectiveness of the adaptation mechanism.  Three key assumptions
enter flow control, Eq.~(\ref{a_i_flow_introduction}):

\begin{itemize}

  \item Represented by $x_{{\rm r},i}(t)$, we assume that there is
  a physical separation between the recurrent input that a neuron
  receives and its external inputs. This is necessary because
  $x_{{\rm r},i}(t)$ is explicitly used in the update rule
  of the synaptic scaling factors.

  \item Synaptic scaling only affects the weights of recurrent connections.
  However, this assumption is not crucial for the effectiveness
  of our plasticity rule, as we were mostly concerned with
  achieving a preset spectral radius for the recurrent weight
  matrix. If instead the scaling factors acted on both the recurrent
  and external inputs, this would lead to an ``effective" external
  input $I'_i(t) := a_i I_i(t)$. However, $a_i$ only affecting
  the recurrent input facilitated the parameterization of
  the external input by means of its variance,
  see Section~\ref{sect:XOR}, a choice of convenience.

  \item For (\ref{a_i_flow_introduction}) to function, neurons
  need to able to represent and store squared neural activities.

\end{itemize}

Whether these three preconditions are satisfied by
biological neurons needs to be addressed in future
studies.

\bigskip

\subsubsection{Variance control}

The second adaptation rule, variance control, has the form
%
  \begin{align}
  \label{a_i_variance}
  a_i(t) &= a_i(t-1) + \epsilon_{\rm a} \left[
  \sigma_{{\rm t},i}^2(t) -
  {\left( y_i(t) - \mu^{\rm y}_i(t) \right)}^2\right] \\
  \label{sigm_target}
  \sigma_{{\rm t},i}^2(t) &= 1 - \frac{1}{\sqrt{1 + 2R_{\rm t}^2
      y_i(t)^2
      +
      2\sigma_{{\rm ext},i}^2(t)}}  
  \; .
  \end{align}
%
Eq.~(\ref{a_i_variance}) drives the average variance of each
neuron towards a desired target variance $\sigma_{{\rm t},i}^2(t)$
at an adaptation rate $\epsilon_{\rm a}$ by calculating the momentary 
squared difference between the local activity $y_i(t)$ and its trailing 
average $\mu^{\rm y}_i(t)$. Eq.~(\ref{sigm_target}) calculates the 
target variance as a function of the target spectral radius $R_{\rm t}$, 
the current local square activity $y^2_i(t)$ and a trailing average 
$\sigma^2_{{\rm ext},i}(t)$ of the local variance of the external 
input signal. When all $a_i(t)$ reach a steady state, 
the average neural variance equals the target given by 
(\ref{sigm_target}). According to a mean-field approach that is described 
in Section~\ref{sect:MF_theory}, reaching this state then results in 
a spectral radius $R_{\rm a}$ that is equal to the target $R_{\rm t}$ 
entering (\ref{sigm_target}). Intuitively, it is to be  expected that 
$\sigma_{{\rm t},i}^2$ is a function of both the spectral radius
and the external driving variance: The amount of fluctuations in the 
network activity is determined by the dynamic interplay between the strength 
of the external input as well as the recurrent coupling.

A full description of the auxiliary equations and variables used to 
calculate $\mu^{\rm y}_i(t)$ and $\sigma^2_{{\rm ext},i}(t)$ can be 
found in Section~\ref{sect:model}.

Similar to flow control, we also considered a non-local version
for comparative reasons, where (\ref{sigm_target}) is replaced with
%
  \begin{equation}
  \label{sigm_target_global}
  \sigma_{{\rm t},i}^2(t) = 1 - \frac{1}{\sqrt{1 + 2R_{\rm t}^2
    ||\mathbf{y}(t)||^2/N
    +
    2\sigma_{{\rm ext},i}^2(t)}} \; .
  \end{equation}
%
Again, $||\cdot||$ denotes the euclidean norm.
Before proceeding to the results, we discuss the mathematical
background of the proposed adaptation rules in some detail.

\bigskip

\subsection{Autonomous spectral radius regulation}
\label{sect_specrad_reg}

There are some interesting aspects to the theoretical
framework at the basis of the here proposed regulatory
scaling mechanisms.

The circular law of random matrix theory states that
the eigenvalues $\lambda_j$ are distributed uniformly on
the complex unit disc if the elements of a real
$N\times N$
matrix are drawn from distributions having zero mean and
standard deviation $1/\sqrt{N}$ \citep{tao2008random}.
Given that the internal weight matrix $\widehat{W}$ ( $\widehat{\cdot}$~denoting matrices)
with entries $W_{ij}$ has
$p_{\rm r} N$ non-zero elements per row ($p_{\rm r}$
is the connection probability), the circular law
implies that the spectral radius of $a_i W_{ij}$, the maximum of
$|\lambda_j|$, is unity when the synaptic scaling factors
$a_i$ are set uniformly to $1/\sigma_{\rm w}$, where $\sigma_{{\rm w}}$
is the standard deviation of $W_{ij}$.
Our goal is to investigate adaptation rules for the synaptic
scaling factors that are based on dynamic quantities,
which includes the membrane potential $x_i$, the neural
activity $y_i$ and the input $I_i$.

The circular law, i.\ e.\ a $N \times N$ matrix with i.i.d.\
entries with zero mean and $1/N$ variance approaching a spectral
radius of one as $N \rightarrow \infty$, can be generalized.
\citet{Rajan2006} investigated the case where the statistics
of the columns of the matrix differ in their means and variances:
given row-wise E-I balance for the recurrent weights, the
square of the spectral radius of a random $N\times N$ matrix
whose columns have variances $\sigma^2_i$ is
$N\left\langle \sigma^2_i \right\rangle_i$ for $N \rightarrow \infty$.
Since the eigenvalues are invariant under transposition, 
this result also holds for row-wise variations of variances
and column-wise E-I balance. While the latter is not explicitly
enforced in our case, deviations from this balance are expected to 
tend to zero for large $N$ given the statistical assumptions
that we made about the matrix elements $W_{ij}$.
Therefore, the result can be applied to our model, where node-specific gain
factors $a_i$ are applied to each row of the recurrent
weight matrix. Thus, the spectral radius $R_{\rm a}$ of the 
\emph{effective random matrix} $\widehat{W}_{\rm a}$ 
with entries $a_iW_{ij}$ 
(as entering (\ref{x_r_i_introduction})) is
%
  \begin{equation}
  R_{\rm a}^2 \approxeq \frac{1}{N} \sum_i R_{{\rm a},i}^2,
  \qquad\quad
  R_{{\rm a},i}^2 := a^2_i \sum_j \left(W_{ij}\right)^2\,,
  \label{R_a}
  \end{equation}
%
for large $N$, when assuming that the distribution underlying
the \textit{bare weight matrix} $\widehat{W}$ with entries $W_{ij}$
 has zero mean. Note that $R^2_{\rm a}$ can
be expressed alternatively in terms of the Frobenius norm
$\left\lVert\widehat{W}_{\rm a} \right\rVert_{\rm F}$, via
%
  \begin{equation}
  R_{\rm a}^2 \approxeq \left\lVert\widehat{W}_{\rm a}
  \right\rVert^2_{\rm F} / N \, .
  \end{equation}
%
We numerically tested Eq.~(\ref{R_a}) for $N=500$
and heterogeneous random sets of $a_i$ drawn from a
uniform $[0,1]$-distribution and found a very close
match to the actual spectral radii ($1$-$2\%$ relative
error). Given that the $R_{{\rm a}, i}$ can be
interpreted as per-site estimates for the spectral radius,
one can use the generalized circular law (\ref{R_a}) to
regulate  $R_{\rm a}$ on the basis of local adaptation
rules, one for every $a_i$.

For the case of flow control, the adaptation rule is
derived using a comparison between the variance of
neural activity that is present in the network with
the recurrent contribution to the membrane potential.
A detailed explanation is presented in
Section~\ref{sec:sing_values} and
Section~\ref{sect:flow_theo}.
In short, we propose that
%
  \begin{equation}
  {\big\langle\,{||\mathbf{x}_{\rm r}(t)||}^2\,\big\rangle}_{\rm t} \approxeq
  R_{\rm a}^2\,
  {\big\langle\,{||\mathbf{y}(t-1)||}^2\,\big\rangle}_{\rm t}\,,
  \label{flow_R_a_introduction}
  \end{equation}
%
where $x_{{\rm r},i}$ is the recurrent
contribution to the membrane potential $x_i$. This stationarity
condition leads to the adaptation rule given in
Eq.~(\ref{a_i_flow_introduction}). An analysis of the dynamics of this
adaptation mechanisms can be found in Section \ref{sect:flow_theo}.

Instead of directly imposing Eq.~(\ref{flow_R_a_introduction})
via an appropriate adaptation mechanism, we also considered the
possibility of transferring this condition into a set point for
the variance of neural activities as a function the external
driving. To do so, we used a mean-field approach to describe the
effect of recurrent input onto the resulting neural activity
variance. An in-depth discussion is given in
Section~\ref{sect:MF_theory}. This led to the update rule given by
Eq.~(\ref{a_i_variance}) and (\ref{sigm_target}) for variance control.

\bigskip

\subsection{Testing protocols}
\label{sect:testing_protocols}

We used several types of input protocols for testing the
here proposed adaptation mechanisms, as well as for assessing
the task performance discussed in Section \ref{sect:XOR}. The
first two variants concern distinct biological scenarios:

\begin{itemize}
  \item {\em Binary.}
  Binary input sequences correspond to a situation when
  a neural ensemble receives input dominantly from a
  singular source, which itself has only two states,
  being either active or inactive. Using binary input
  sequences during learning is furthermore consistent
  with the non-linear performance test considered here for
  the echo-state network as a whole, the delayed XOR-task.
  See Section~\ref{sect:XOR}. For symmetric binary inputs,
  as used, the source signal $u(t)$ is drawn from
  $\pm1$.

  \item {\em Gaussian.}
  Alternatively one can consider the situation that
  a large number of essentially uncorrelated input
  streams are integrated. This implies random Gaussian
  inputs signals. Neurons receive in this case zero-mean
  independent Gaussian noise.
\end{itemize}

Another categorical dimension concerns the distribution
of the afferent synaptic weights. Do all neurons receive
inputs with the same strength, or not? As a
quantifier for the individual external input strengths,
 the variances
$\sigma^2_{{\rm ext}, i}$ of the local external
input currents where taken into account. We distinguished two cases

\begin{itemize}
  \item {\em Heterogeneous.}
  In the first case, the $\sigma^2_{{\rm ext}, i}$ are
  quenched random variables. This means that each neuron
  is assigned a random value $\sigma^2_{{\rm ext}, i}$
  before the start of the simulation, as drawn from a
  half-normal distribution parameterized by
  $\sigma = \sigma_{{\rm ext}}$. This ensures that
  the expected average variance
  $\big\langle \sigma^2_{{\rm ext}, i} \big\rangle$
  is given by $\sigma^2_{{\rm ext}}$.

  \item {\em Homogeneous.}
  In the second case, all $\sigma^2_{{\rm ext}, i}$ are
  assigned the identical global value $\sigma^2_{{\rm ext}}$.
\end{itemize}


Overall, pairing
``binary'' vs.\, ``Gaussian'' and
``heterogeneous'' vs.\, ``homogeneous'',
leads to a total of four different input protocols,
i.\,e.\
``heterogeneous binary'',
``homogeneous binary'',
``heterogeneous Gaussian'' and
``homogeneous Gaussian''.
If not otherwise stated, numerical simulations were done
using networks with $N=500$ sites and a connection probability
$p_{\rm r}=0.1$.

\bigskip

\subsection{Performance testing of flow control}
\label{sect:flow_control_results}

\begin{figure}[t]
\includegraphics[width=1.0\textwidth] 
{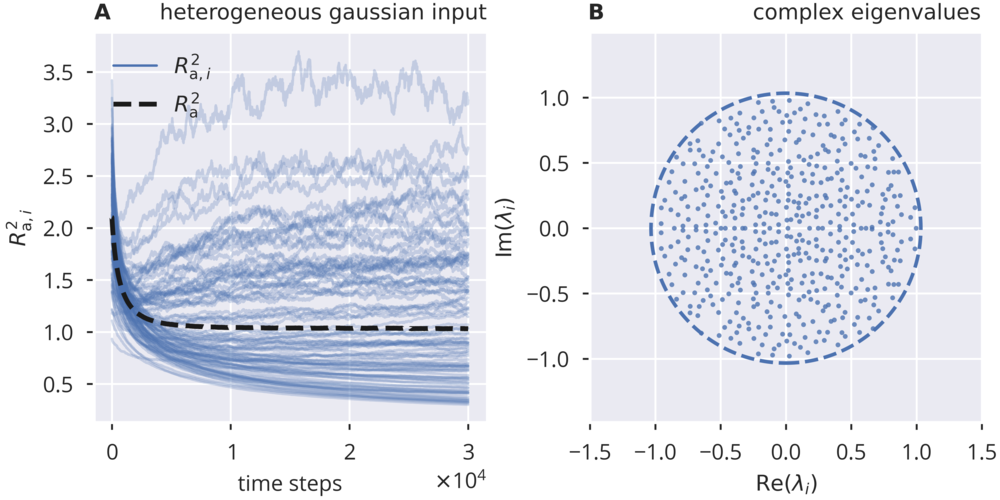}
\caption{{\bf Online spectral radius regulation using flow control.}
The spectral radius $R_{\rm a}$ and the respective local estimates
$R_{{\rm a},i}$ as defined by (\ref{R_a}). For the input protocol 
see Section~\ref{sect_input}.
{\bf A}: Dynamics of $R^2_{{\rm a},i}$ and $R^2_{\rm a}$,
in the presence of heterogeneous independent Gaussian inputs.
Local adaptation.
{\bf B}: Distribution of eigenvalues of the corresponding
effective synaptic matrix $\widehat{W}_{\rm a}$,
after adaptation. The circle denotes the spectral radius.
}
\label{fig_R_a_regulation}
\end{figure}

In Figure~\ref{fig_R_a_regulation}, we present a simulation
using flow control for heterogeneous Gaussian input with
an adaptation rate $\epsilon_{\rm a}=10^{-3}$. The
standard deviation of the external driving was set to
$\sigma_{\rm ext}=0.5$. The spectral radius of $R_{\rm a}$
of $\widehat{W}_{\rm a}$ was tuned to the target $R_{\rm t} = 1$
with high precision, even though the local, row-wise estimates
$R_{{\rm a},i}$ showed substantial deviations from the target.

\begin{figure}[htbp]
  \begin{center}
    \includegraphics[width=\textwidth]
    {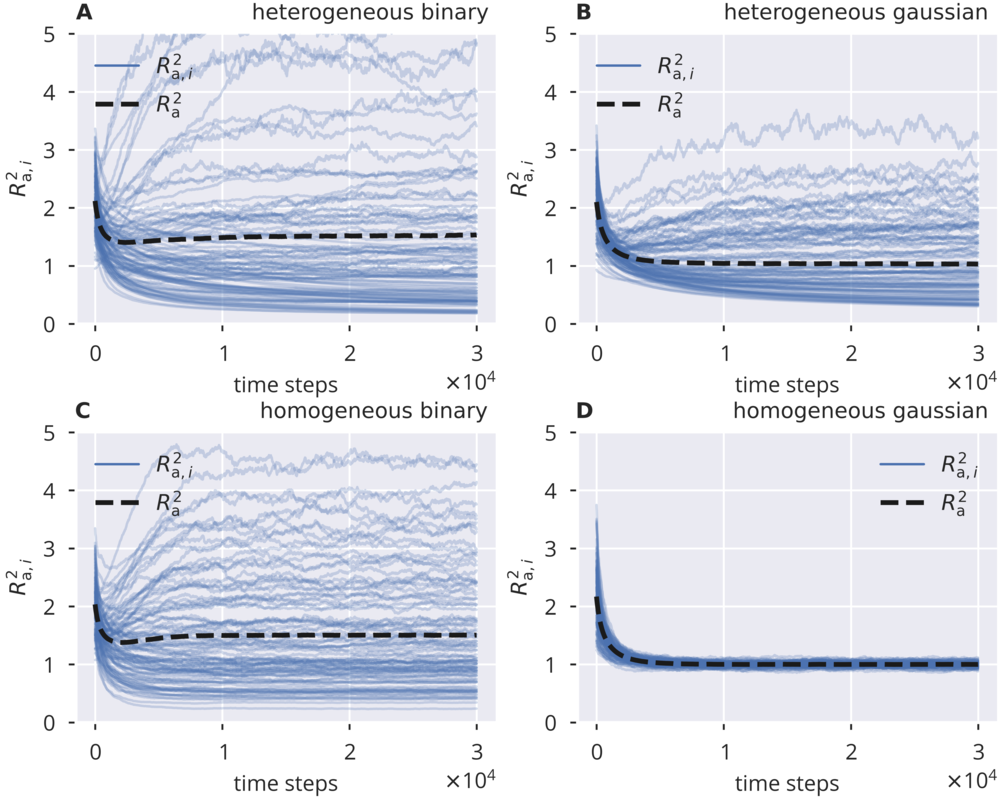}
  \end{center}
  \caption{{\bf Adaptation dynamics, flow control, local} 
  Panels {\bf A}--{\bf D} show the dynamics of the square spectral 
  radius $R^2_{\rm a}$ and local estimates $R^2_{{\rm a},i}$ 
  under local flow control for different 
  input protocols, as given in the panel titles.
  }
  \label{S_flow_control_local_Fig}
\end{figure}

\begin{figure}[htbp]
  \begin{center}
    \includegraphics[width=1.0\textwidth]
    {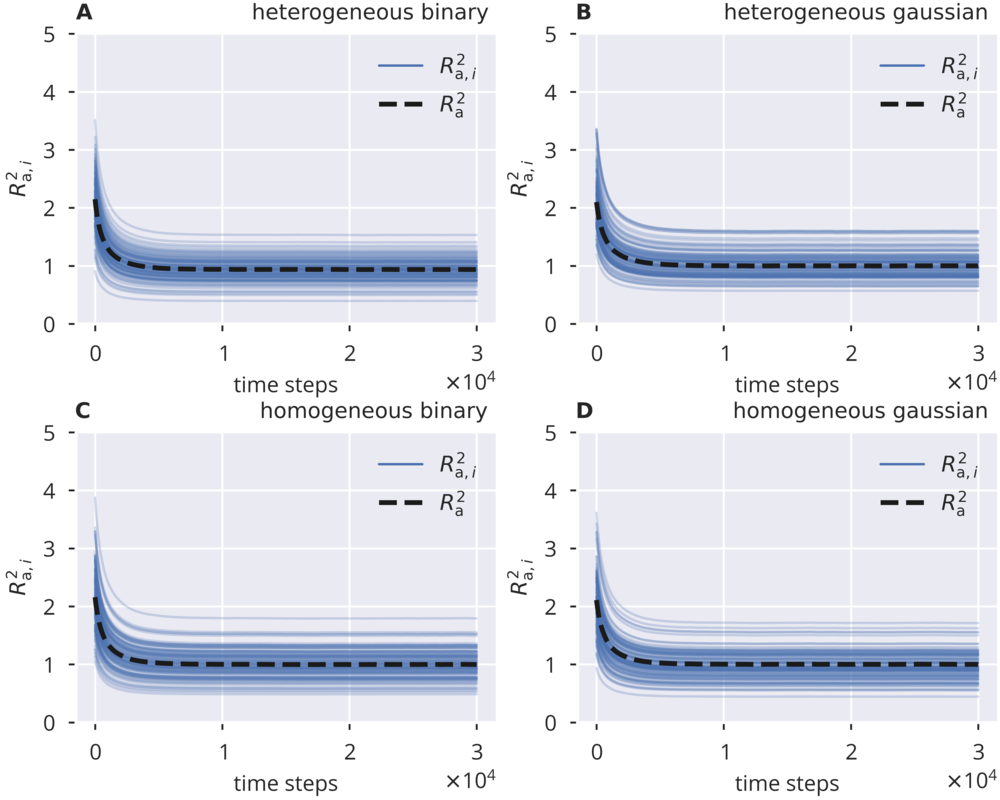}
  \end{center}
  \caption{{\bf Adaptation dynamics, flow control, global} 
    Panels {\bf A}--{\bf D} show the dynamics of the square spectral 
    radius $R^2_{\rm a}$ and local estimates $R^2_{{\rm a},i}$ 
    under global flow control for different 
    input protocols, as given in the panel titles.
  }
\label{S_flow_control_global_Fig}
\end{figure}

We further tested the adaptation with other input protocols, see
Section~\ref{sect:testing_protocols} and Figure~\ref{S_flow_control_local_Fig}.
We found that flow control 
robustly led to the desired spectral
radius $R_{\rm t}$ under all Gaussian input protocols,
while binary input caused $R_{\rm a}$ to converge to higher
values than $R_{\rm t}$. However, when using global adaptation,
as given by Eq.~(\ref{delta_R_global_introduction}),
all input protocols resulted in the correctly tuned spectral radius,
see Figure~\ref{S_flow_control_global_Fig}.


Numerically, we found that the time needed to converge 
to the stationary states depended substantially on 
$R_{\rm t}$, slowing down when the spectral radius becomes
small. It is then advantageous, as we have done, to scale the
adaptation rate $\epsilon_{\rm a}$ inversely with the
trailing average $\bar{x}_{\rm r}^2$ of $||x_{\rm r}||^2$,
viz as
$\epsilon_{\rm a} \to \epsilon_{\rm a}/\bar{x}_{\rm r}^2$. 
An exemplary plot showing the effect of this scaling is shown in 
Fig.~\ref{fig:flow_renorm}, see 
Section~\ref{sect:renorm_flow_control} for further details.

  To evaluate the amount of deviation from the target spectral
  radius under different input strengths and protocols, we plotted
  the difference between the resulting spectral radius and the target 
  spectral radius for a range of external input strength, quantified 
  by their  standard deviation $\sigma_{{\rm ext}}$. Results for 
  different input protocols are shown in Figure~\ref{S1_Fig} in the 
  supplementary material. For correlated binary input, increasing 
  the input strength resulted in stronger deviations from
  the target spectral radius. On the other hand, uncorrelated
  Gaussian inputs resulted in perfect alignment for the entire
  range of input strengths that we tested.


\bigskip
\bigskip

\subsection{Perfomance testing of variance control}
\label{sect:variance_control_results}
  
\begin{figure}[t]
\includegraphics[width=1.0\textwidth]
{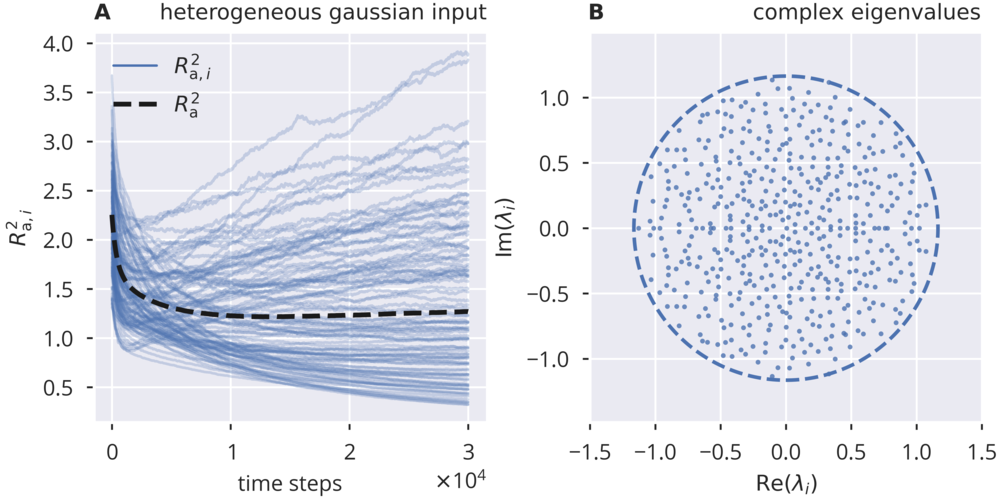}
\caption{{\bf Online spectral radius regulation using variance control.}
The spectral radius $R_{\rm a}$ and the respective local estimates
$R_{{\rm a},i}$ as defined by (\ref{R_a}). For the input protocol 
see Section~\ref{sect_input}.
{\bf A}: Dynamics of $R^2_{{\rm a},i}$ and $R^2_{\rm a}$,
in the presence of heterogeneous independent
Gaussian inputs. Local adaptation.
{\bf B}: Distribution of eigenvalues of the corresponding
effective synaptic matrix $\widehat{W}_{\rm a}$.
The circles denote the respective spectral radius.
}
\label{fig_R_a_regulation_var}
\end{figure}

\begin{figure}[htbp]
  \begin{center}
    \includegraphics[width=\textwidth]
    {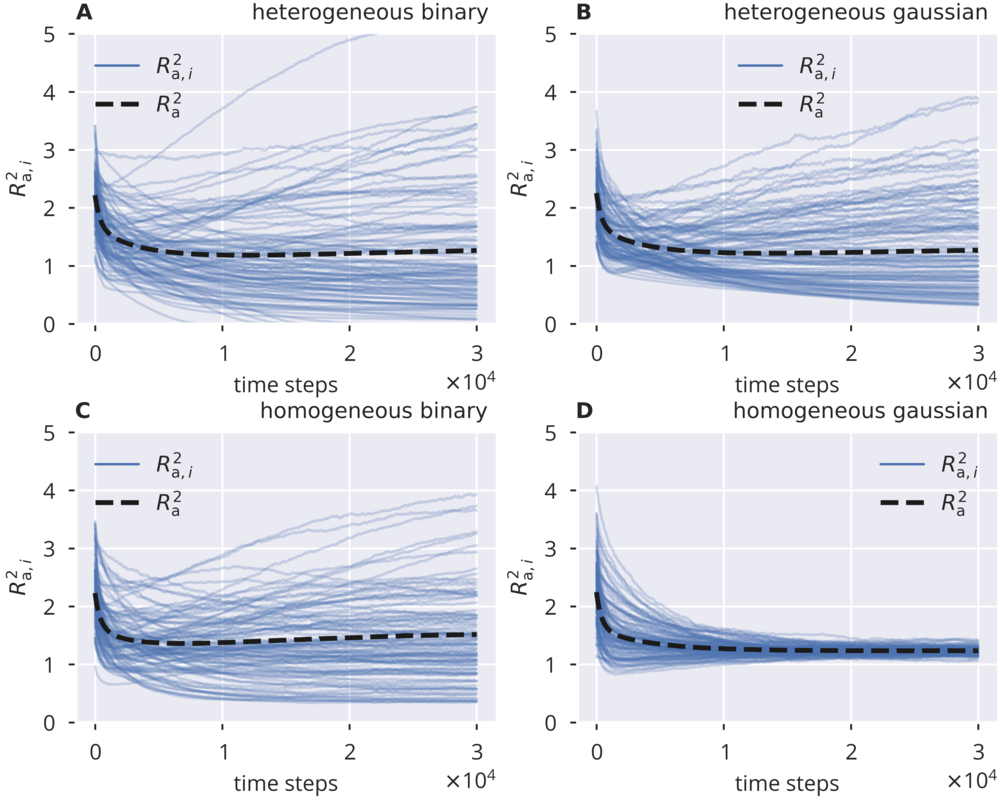}
  \end{center}
  \caption{{\bf Adaptation dynamics, variance control, local} 
    Panels {\bf A}--{\bf D} show the dynamics of the square spectral 
    radius $R^2_{\rm a}$ and local estimates $R^2_{{\rm a},i}$ 
    under local variance control for different 
    input protocols, as given in the panel titles.
  }
\label{S_var_control_local_Fig}
\end{figure}

\begin{figure}[htbp]
  \begin{center}
    \includegraphics[width=1.0\textwidth]
    {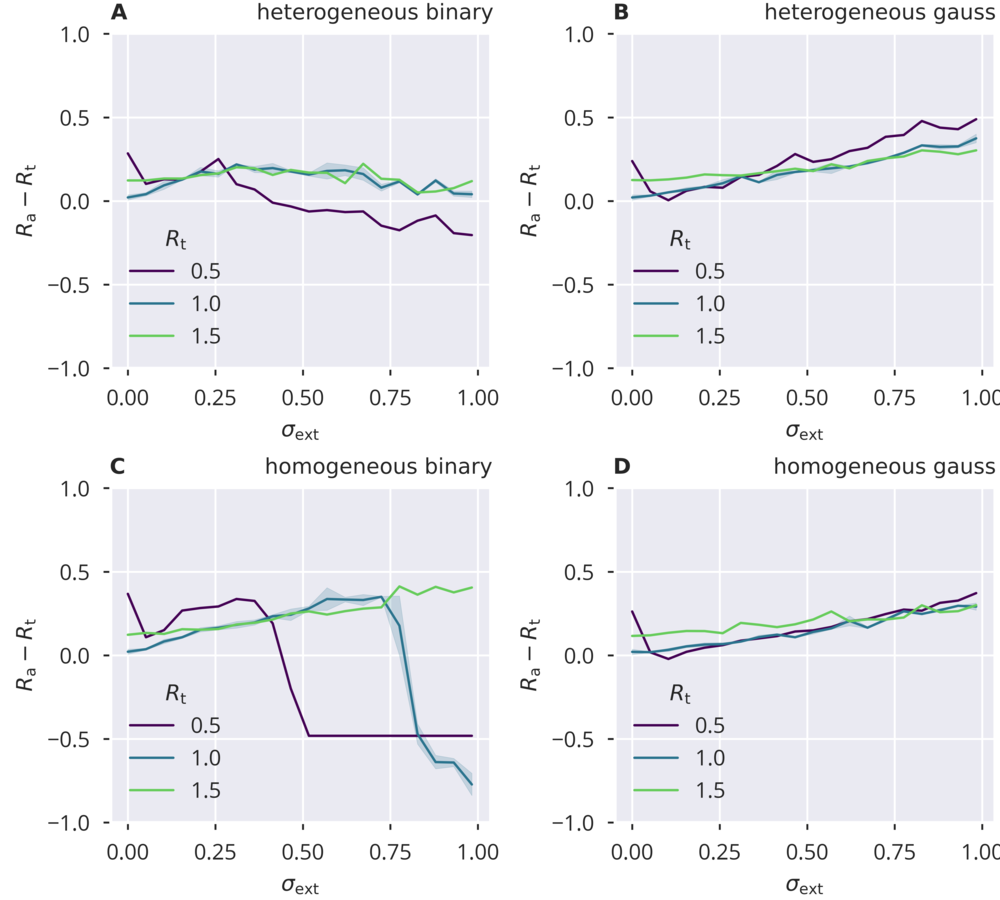}
  \end{center}
    \caption{{\bf Difference of the spectral radius after adaptation and the 
        target spectral radius, variance control} For different 
      standard deviations $\sigma_{{\rm ext}}$ of the external 
      input, the error $R_{\rm a} - R_{\rm t}$ between the 
      resulting spectral radius $R_{\rm a}$ 
      and the target spectral radius $R_{\rm t}$was 
      determined. Heterogeneous binary input 
      ({\bf A}) led to a good alignment. On the other hand, 
      homogeneous binary input ({\bf C}) caused strong 
      deviation from the target for larger input. 
      Heterogeneous/homogeneous Gaussian input ({\bf B}/{\bf D}) 
      both resulted in positive deviations that 
      increased for larger input strengths. 
      Local adaptation was used for both panels.}
\label{S2_Fig}
\end{figure}

In comparison, variance control, shown in 
Figure~\ref{fig_R_a_regulation_var} and 
Figure~\ref{S_var_control_local_Fig}, resulted in 
notable deviations from $R_{\rm t}$, for both 
uncorrelated Gaussian and correlated binary input. 
As for flow control, we also calculated the deviations from 
$R_{\rm t}$ as a function of $\sigma_{{\rm ext}}$, see Figure~\ref{S2_Fig}. 
For heterogeneous binary input, deviations from the target spectral 
radius did not increase monotonically as a function of the 
input strength (Figure~\ref{S2_Fig}A), reaching a peak 
at $\sigma_{{\rm ext}}\approx 0.4$ for target spectral 
radii larger than $1$. For homogeneous binary input, 
we observed a substantial negative mismatch of the spectral 
radius for strong external inputs, see Figure~\ref{S2_Fig}C.

\begin{figure}[htbp]
  \begin{center}
    \includegraphics[width=1.0\textwidth]
    {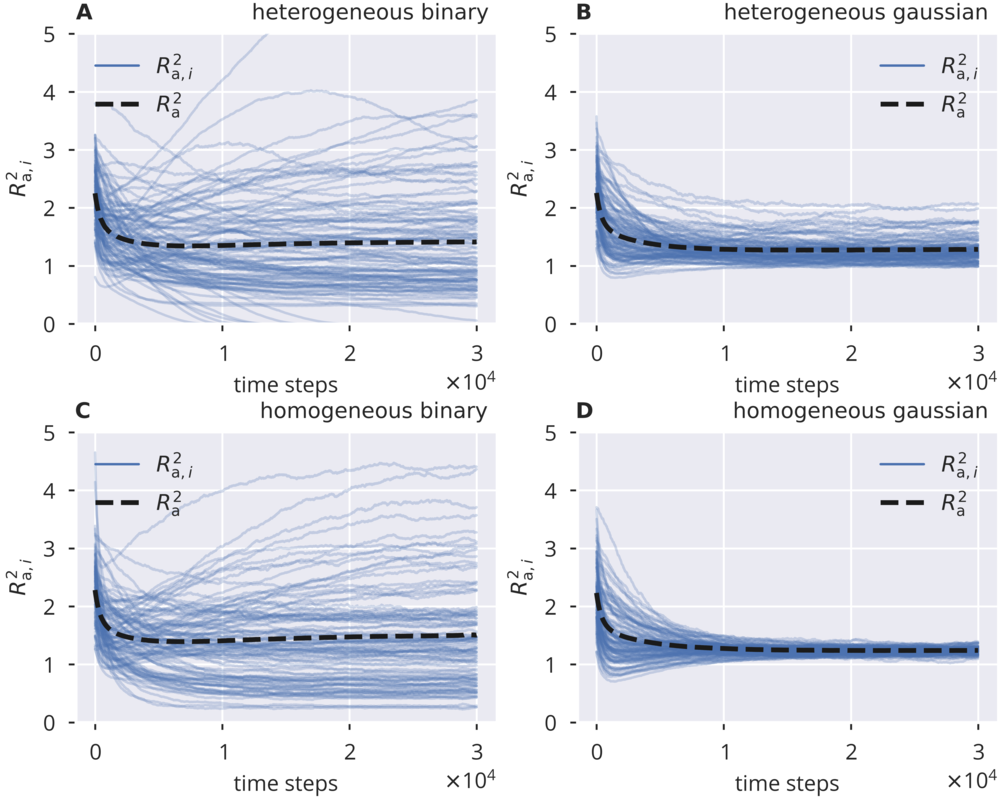}
  \end{center}
  \caption{{\bf Adaptation dynamics, variance control, global} 
    Panels {\bf A}--{\bf D} show the dynamics of the square spectral 
    radius $R^2_{\rm a}$ and local estimates $R^2_{{\rm a},i}$ 
    under global variance control for different 
    input protocols, as given in the panel titles.
  }
\label{S_var_control_global_Fig}
\end{figure}

\begin{figure}[htbp]
  \begin{center}
    \includegraphics[width=1.0\textwidth]
    {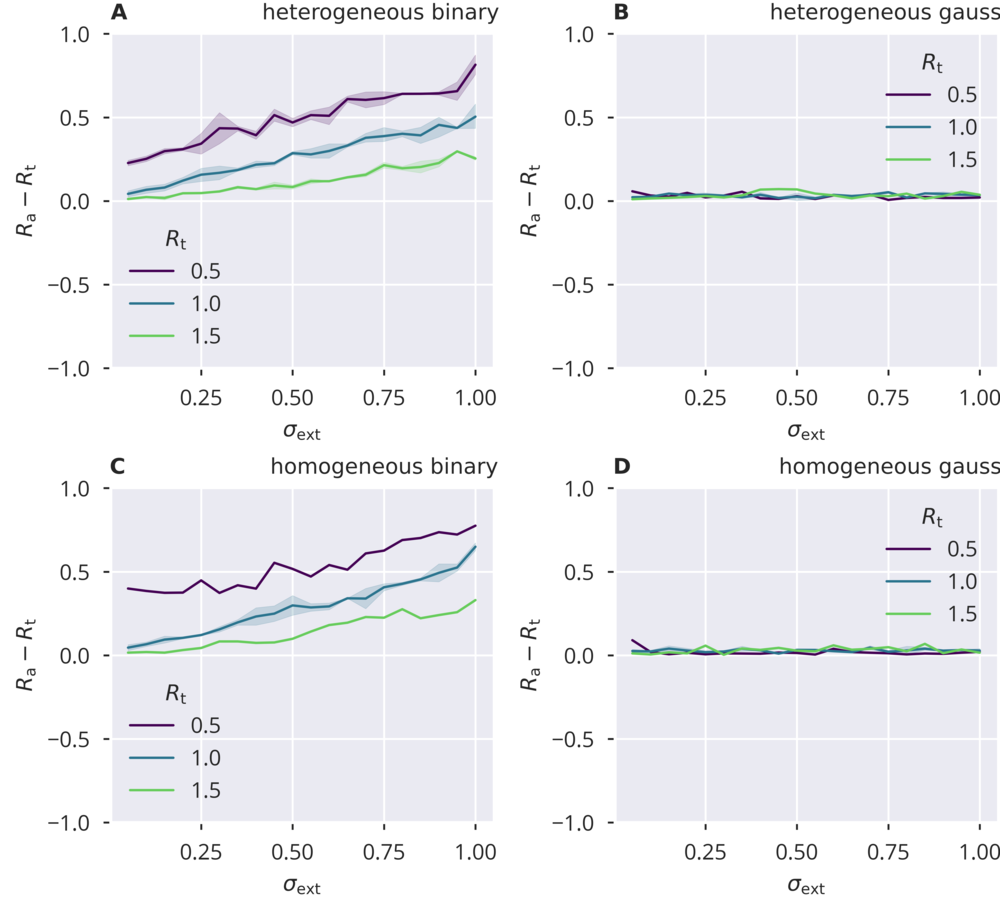}
  \end{center}
  \caption{{\bf Difference of the spectral radius after adaptation and the 
    target spectral radius, flow control} For different 
    standard deviations $\sigma_{{\rm ext}}$ of the external 
    input, the error $R_{\rm a} - R_{\rm t}$ between the 
    resulting spectral radius $R_{\rm a}$ 
    and the target spectral radius $R_{\rm t}$was 
    determined. Heterogeneous/homogeneous binary input 
    ({\bf A}/{\bf C}) led to positive deviations from the 
    target spectral radius for stronger external input. 
    Heterogeneous/homogeneous Gaussian input ({\bf B}/{\bf D}) 
    yielded perfect alignment between $R_{\rm a}$ and 
    $R_{\rm t}$. Local adaptation was used for both panels.}
  \label{S1_Fig}
\end{figure}

Overall, we found that variance control did not exhibit
the same level of consistency in tuning the system towards a
desired spectral radius, even though it did perform better in
some particular cases (compare Figure~\ref{S1_Fig}A
for large $\sigma_{{\rm ext}}$ with Figure~\ref{S2_Fig}).
Moreover, variance control exhibited deviations from the
target (shown in Figure~\ref{S_var_control_global_Fig}) 
even when a global adaptation rule was used, as 
defined in (\ref{sigm_target_global}).
This is in contrast to the global variant of flow control, which,
as stated in the previous section, robustly tuned the spectral 
radius to the desired value even in the presence of strongly
correlated inputs.

\bigskip

\subsection{Spectral radius, singular values and global Lyapunov exponents}
\label{sec:sing_values}

Apart from the spectral radius $R_{\rm a}$ of the matrix
$\widehat{W}_{\rm a}$, one may consider the relation between the
adaptation dynamics and the respective singular values
$\sigma_i$ of $\widehat{W}_{\rm a}$. We recall
that the spectrum of
$\hat{U}_{\rm a}=\widehat{W}_{\rm a}^\dagger
\widehat{W}_{\rm a}$
is given by the squared singular values, $\sigma_i^2$,
and that the relation $||\mathbf{x}_{\rm r}||^2 =
\mathbf{y}^\dagger \widehat{W}_{\rm
  a}^\dagger\widehat{W}_{\rm a}
\mathbf{y}$ holds. Now, assume that the time-averaged projection
of neural activity $\mathbf{y}=\mathbf{y}(t)$ onto
all eigenvectors of $\hat{U}_{\rm a}$ is approximately the same, that is,
there is no preferred direction of neural activity in phase
space. From this idealized case, it follows that the time
average of the recurrent contribution to the
membrane potential can be expressed with
%
  \begin{equation}
  {\big\langle\,{||\mathbf{x}_{\rm r}||}^2\,\big\rangle}_{\rm
    t}
  \approx \frac{{\big\langle\,||\mathbf{y}||^2\,\big\rangle}_{\rm
      t} }{N} \sum_i \sigma_i^2
  = \frac{{\big\langle\,||\mathbf{y}||^2,\big\rangle}_{\rm
      t} }{N} \sum_{i,j}
  {\big(a_iW_{ij}\big)}^2
  \label{SVD_x_r}
  \end{equation}
%
as the rescaled average of the $\sigma_i^2$. For
the second equation, we used the fact that the
$\sum_i\sigma_i^2$ equals the sum of all matrix
elements squared
\citep{sengupta1999distributions,shen2001singular}.
With (\ref{R_a}), one finds that (\ref{SVD_x_r})
is equivalent to
${{\big\langle\,||\mathbf{x}_{\rm r}||}^2,\big\rangle}_{\rm
  t} = R_a^2 {\big\langle\,||\mathbf{y}||^2,\big\rangle}_{\rm
  t}$ and hence to the flow condition
(\ref{flow_R_a_introduction}). This result can be generalized,
as done in Section~\ref{sect:flow_theo},
to the case that the neural activities have inhomogeneous
variances, while still being uncorrelated with zero mean.
We have thus shown that the stationarity condition leads 
to a spectral radius of (approximately) unity.

It is worthwhile to note that the singular values of
$\hat{U}_{\rm a}=\widehat{W}_{\rm a}^\dagger
\widehat{W}_{\rm a}$ do exceed unity when $R_{\rm a} = 1$.
More precisely, for a random matrix with i.i.d.\ entries,
one finds in the limit of large $N$ that the largest singular
value is given by $\sigma_{\rm max} = 2 R_{\rm a}$,
in accordance with the Marchenko-Pastur law for 
random matrices \citep{Marcenko1967}.
Consequently, directions in phase space exist
in which the norm of the phase space vector is elongated
by factors greater than one. Still, this does not contradict the fact
that a unit spectral radius coincides with the transition
to chaos for the non-driven case. The reason is that
the global Lyapunov exponents are given by
%
  \begin{equation}
  \lim\limits_{n\rightarrow \infty}
  \frac{1}{2n}\ln\left(\left(\widehat{W}_{\rm
    a}^n\right)^\dagger
  \widehat{W}_{\rm a}^n \right)
  \end{equation}
%
which eventually converge to $\ln \lVert \lambda_i \rVert$,
see Figure~\ref{S3_Fig} in the supplementary material and
\citet{wernecke2019chaos}, where
$\lambda_i$ is the $i$th eigenvalue of $\widehat{W}_{\rm a}$.
The largest singular value of the $n$th power of a random
matrix with a spectral radius $R_{\rm a}$ scales like
$R_{\rm a}^{n}$ in the limit of large powers $n$.
The global Lyapunov exponent goes to zero as a consequence
when $R_{\rm a}\to1$.

\medskip

\subsection{XOR-memory recall}
\label{sect:XOR}

To this point, we presented results regarding the
effectiveness of the introduced adaptation rules.
However, we did not account for their
effects onto a given learning task. Therefore,
we tested the performance of locally adapted networks under
the delayed XOR task, which evaluates the memory capacity
of the echo state network in combination with a
non-linear operation. For the task, the XOR operation
is to be taken with respect to a delayed pair of two
consecutive binary inputs signals, $u(t\!-\!\tau)$ and
$u(t\!-\!\tau\!-\!1)$, where $\tau$ is a fixed time delay. 
The readout layer is given by a single unit, which has 
the task of reproducing
%
  \begin{equation}
  \label{XOR_f_t}
  f_\tau(t) =
  \mathrm{XOR}\left[u(t\!-\!\tau),u(t\!-\!\tau\!-\!1)\right],
  \qquad\quad
  t,\tau=1,\,2,\, \dots\,,
  \end{equation}
%
where $\mathrm{XOR}[u,u']$ is $0/1$ if $u$ and $u'$
are identical/not identical.

The readout vector $\mathbf{w}_{\rm out}$ is trained
with respect to the mean squared output error,
%
  \begin{equation}
  {||\widehat{Y} \mathbf{w}_{\rm out} - \mathbf{f}_\tau||}^2
  + \alpha {|| \mathbf{w}_{\rm out} ||}^2\,,
  \label{XOR_error}
  \end{equation}
%
using ridge regression on a sample batch of
$T_{\rm batch} = 10 N$ time steps, here for $N=500$,
and a regularization factor of $\alpha=0.01$.
The batch matrix $\widehat{Y}$, of size
$T_{\rm batch} \times (N+1)$, holds the neural
activities as well as one node with constant
activity serving as a potential bias.
Similarly, the readout (column) vector $\mathbf{w}_{\rm out}$
is of size $(N+1)$. The $T_{\rm batch}$
entries of $\mathbf{f}_\tau$ are the $f_\tau(t)$, viz the
target values of the XOR problem.
Minimizing (\ref{XOR_error}) leads to
%
  \begin{equation}
  \mathbf{w}_{\rm out} = {\left(\widehat{Y}^\dagger
    \widehat{Y}
    + \alpha^2 \hat{\mathbb{1}} \right)}^{-1} \widehat{Y}^\dagger\,
  \mathbf{f}_\tau \,.
  \end{equation}
%
The learning procedure was repeated independently
for each time delay $\tau$. We quantified the
performance by the total memory capacity,
$\mathrm{MC}_{\rm XOR}$, as
%
  \begin{align}
  \mathrm{MC}_{\rm XOR} &= \sum_{k=1}^\infty
  \mathrm{MC}_{{\rm XOR},k}
  \label{MC_XOR}
  \\
  \mathrm{MC}_{{\rm XOR},k} &=
  \frac{\mathrm{Cov}^2\left[f_k(t),y_{\rm out}(t)\right]_t}
  {\mathrm{Var}
    \left[f_k(t)\right]_t
    \mathrm{Var}\left[y_{\rm out}(t)\right]_t}\,.
  \label{MC_XOR_k}
  \end{align}
%

This is a simple extension of the usual definition 
of short term memory in the echo state literature \citep{Jaeger2002}.
The activity
$y_{\rm out}=\sum_{i=1}^{N+1} w_{{\rm out},i}\, y_i$
of the readout unit is compared in (\ref{MC_XOR_k})
with the XOR prediction task, with the additional neuron,
$y_{N+1}=1$, corresponding to the bias of the readout
unit. Depending on the mean level of the target signal,
this offset might actually be unnecessary. However,
since it is a standard practice to use an intercept variable
in linear regression models, we decided to include it
into the readout variable $y_{\rm out}$. The variance
and covariance are calculated with respect to the
batch size $T_{\rm batch}$.

\begin{figure}[t]
    \includegraphics[width=1.0\textwidth]
    {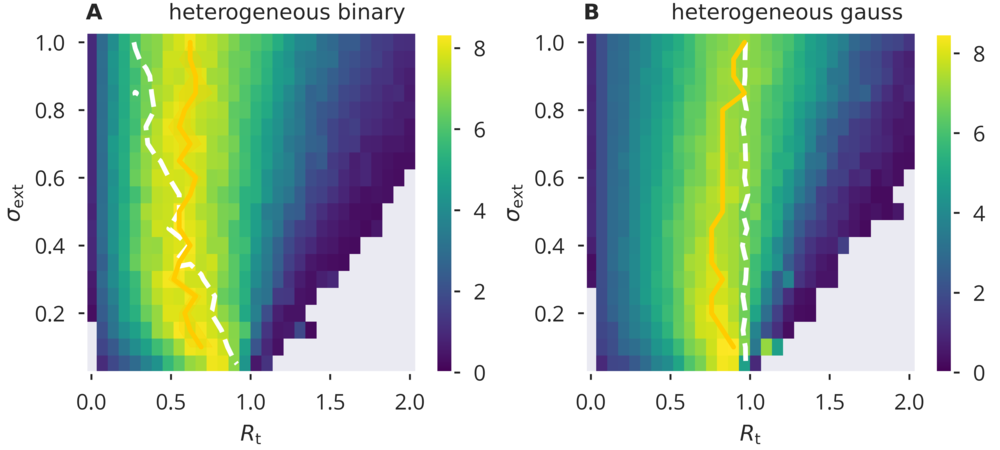}
    \caption{{\bf XOR performance for flow  control.}
Color-coded performance sweeps for the XOR-performance
      (\ref{MC_XOR}) after adaptation using flow control.
      Averaged over five trials. The input
      has variance $\sigma_{\rm ext}^2$ and
      the target for the spectral radius is $R_{\rm t}$.
      A/B panels are for heterogeneous binary/Gaussian
      input protocols. Optimal performance for a given $\sigma_{\rm ext}$ was
      estimated as a trial average (yellow solid
      line) and found to be generally close to criticality, $R_{\rm a} = 1$,
      as measured (white dashed lines).}
    \label{fig:het_performance_sweep_flow_composite}
  \end{figure}

The results for flow control presented in
Figure~\ref{fig:het_performance_sweep_flow_composite}
correspond to two input protocols, heterogeneous
Gaussian and binary inputs. Shown are sweeps over a range
of $\sigma_{\rm ext}$ and $R_{\rm t}$. The update rule
(\ref{a_i_flow_introduction}) was applied to the network for
each pair of parameters until the $a_i$ values converged to a stable
configuration. We then measured the task performance as
described above. Note that in the case of Gaussian input, this
protocol was only used during the adaptation phases. Due to the
nature of the XOR task, binary inputs with the corresponding 
variances are to be used during performance testing.
See Figure~\ref{S4_Fig} in the supplementary material
for a performance sweep using the homogeneous
binary and Gaussian input protocol.
Optimal performance was generally
attained around the $R_{\rm a}\approx 1$ line.
A spectral radius $R_{\rm a}$
slightly smaller than unity was optimal when using Gaussian
input, but not for binary input signals.
In this case the measured spectral radius $R_{\rm a}$
deviated linearly from the target $R_{\rm t}$,
with increasing strength of the input, as
parameterized by the standard deviation
$\sigma_{\rm ext}$. Still, the locus of optimal performance 
was essentially independent of the input strength, 
with maximal performance attained roughly at 
$R_{\rm t}\approx0.55$. Note that the
line $R_{\rm a}=1$ joins $R_{\rm t}=1$ in
the limit $\sigma_{\rm ext}\to0$.

\begin{figure}[t]
    \includegraphics[width=1.0\textwidth]
    {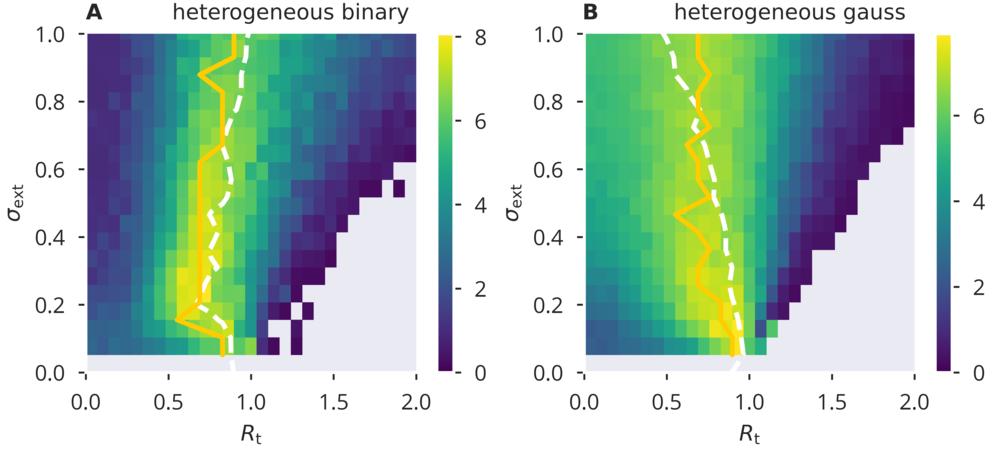}
    \caption{{\bf XOR performance for variance control.}
      Color-coded performance sweeps for the
      XOR-performance
      (\ref{MC_XOR}) after adaptation using variance control.
      Averaged over five trials. The input
      has variance $\sigma_{\rm ext}^2$ and
      the target for the spectral radius $R_{\rm t}$.
      A/B panels are for heterogeneous
      binary/Gaussian
      input protocols. Optimal performance (yellow solid
      line)
      is in general close to criticality, $R_{\rm a} = 1$,
      as measured (white dashed lines).}
    \label{fig:het_performance_sweep_var_composite}
  \end{figure}

Comparing these results to variance control, as shown in
Figure~\ref{fig:het_performance_sweep_var_composite}, we found
that variance control led to an overall lower performance.
To our surprise, for external input with a large variance,
Gaussian input caused stronger deviations from
the desired spectral radius as compared to binary input.
Therefore, in a sense, it appeared to behave opposite to what we
found for flow control. However, similar to flow control, the
value of $R_{\rm t}$ giving optimal performance under a given
$\sigma_{\rm ext}$ remained relatively stable over the range of
external input strength measured. On the other hand, using
homogeneous input, see Figure~\ref{S5_Fig} in the supplementary
material, did cause substantial deviations from the target
spectral radius when using binary input.

  \begin{figure}[t]
    \includegraphics[width=1.0\textwidth]
    {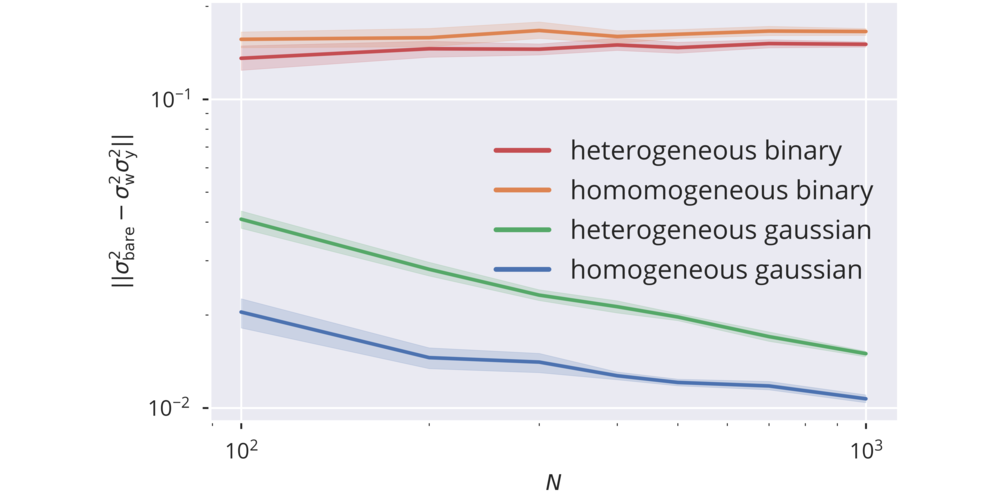}
    \caption{{\bf Size dependence of correlation.}
      Comparison between the variance $\sigma_{\rm
        bare}^2$ of the bare
      recurrent input $x_{\rm bare}=\sum_j W_{ij}y_j$
      with
      $\sigma_{\rm w}^2\sigma_{\rm y}^2$.
      Equality is given when the presynaptic activities
      are statistically
      independent. This can be observed in the limit
      of large network sizes $N$ for uncorrelated input
      data streams
      (homogeneous and heterogeneous Gaussian input
      protocols),
      but not for correlated inputs
      (homogeneous and heterogeneous binary input
      protocols).
      Compare Section~\ref{sect_input} for the input
      protocols.
      Parameters are $\sigma_{\rm ext}\!=\!0.5$,
      $R_{\rm a}\!=\!1$
      and $\mu_{t}\!=\!0.05$.
    }5
    \label{fig:variance_corr_N}
  \end{figure}

  \begin{figure}[t]
    \includegraphics[width=1.0\textwidth]
    {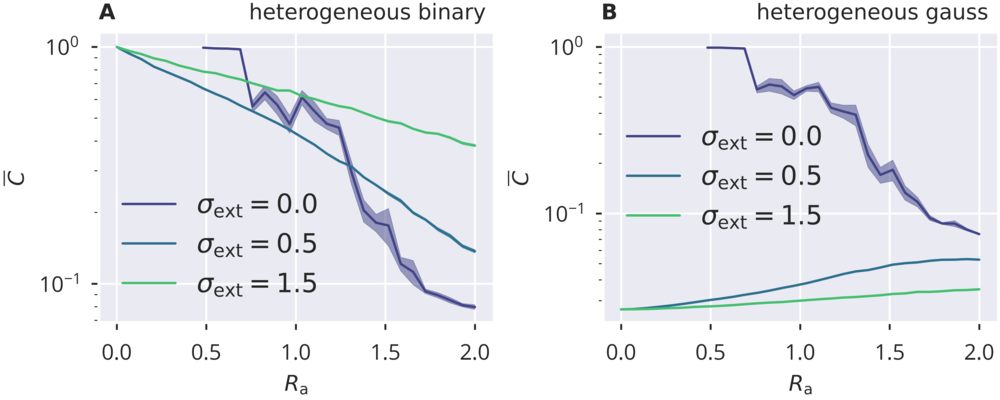}
    \caption{{\bf Input induced activity correlations.}
      For heterogeneous binary and Gaussian inputs
      (A/B), the dependency of mean
      activity
      cross correlations $\bar{C}$, see
      Eq.~(\ref{crossCorr}).
      $\bar{C}$ is
      shown as a function of the spectral radius
      $R_{\rm a}$. Results are obtained
      for $N\!=\!500$ sites by averaging over five trials,
      with shadows indicating the standard error across trials. Correlations
      are
      due to finite-size effect for the autonomous case
      $\sigma_{\rm ext}\!=\!0$.
    }
    \label{fig:het_corr_act_composite}
  \end{figure}

\bigskip

\subsection{Input induced correlations}
\label{sect:correlations}

A crucial assumption leading to the proposed adaptation
rules is the statistical independence of neural activity
for describing the statistical properties of the bare recurrent
contribution to the membrane potential, $x_{\rm bare}=\sum_j W_{ij}y_j$.
In particular, the variance $\sigma^2_{\rm bare}$ of $x_{\rm bare}$
enters the mean-field approach described in Section~\ref{sect:MF_theory}.
Assuming statistical independence across the population for
$y_i(t)$, it is
simply given by $\sigma^2_{\rm bare} = \sigma_{\rm w}^2\sigma_{\rm y}^2$,
where
\begin{equation}
\sigma^2_{\rm w} \equiv \mathrm{Var}\left[\sum_{j=1}^N W_{ij}\right]
\end{equation}
is the variance of the sum of the bare afferent synaptic weights 
(see also Section~\ref{sect:model}).
Being a crucial element of the proposed rules, 
deviations from the prediction of $\sigma^2_{{\rm bare}}$
would also negatively affect the precision of tuning the spectral
radius.
In Figure~\ref{fig:variance_corr_N}, a comparison of the
deviations $|\sigma^2_{{\rm bare}} - \sigma^2_{{\rm w}}\sigma^2_{{\rm y}}|$ 
is presented for the four input protocols introduced in Section~\ref{sect_input}.
For the Gaussian protocols, for which neurons receive
statistically uncorrelated external signals, one observes
that
$\sigma_{\rm bare}^2\to\sigma_{\rm w}^2\sigma_{\rm y}^2$
in the thermodynamic limit $N\to\infty$ via a power law, which
is to be expected when the presynaptic neural activities are
decorrelated.
On the other side, binary $0/1$ inputs act synchronous on
all sites, either with site-dependent or site-independent
strengths (heterogeneous/homogeneous). Corresponding activity
correlations are induced and a finite and only weakly size-dependent
difference between $\sigma_{\rm bare}^2$ and
$\sigma_{\rm w}^2\sigma_{\rm y}^2$
shows up. Substantial corrections to the analytic theory
are to be expected in this case. To this extend we measured
the cross-correlation $C(y_i,y_j)$, defined as
%
  \begin{align}
  \label{crossCorr}
  \bar{C} = \frac{1}{N(N-1)}\sum_{i\ne j} |C(y_i,y_j)|,
  \quad\quad
  C(y_i,y_j) = \frac{\mathrm{Cov}(y_i,y_j)}
  {\sqrt{\mathrm{Cov}(y_i,y_i)\mathrm{Cov}(y_j,y_j)}}\,,
  \end{align}
%
with the covariance given by $\mathrm{Cov}(y_i,y_j) =
\langle (y_i - \langle y_i \rangle_t )
(y_j - \langle y_j \rangle_t ) \rangle_t$.
For a system of $N=500$ neurons the results for the
averaged absolute correlation $\bar{C}$ are presented in
Figure~\ref{fig:het_corr_act_composite}
(see Figure~\ref{S6_Fig} in the supplementary material
for homogeneous input protocols). Autonomous
echo-state layers are in chaotic states when supporting
a finite activity level, which implies that correlations
vanish in the thermodynamic limit $N\to\infty$. The
case $\sigma_{\rm ext}=0$, as included in
Figure~\ref{fig:het_corr_act_composite}, serves
consequently as a yardstick for the magnitude of
correlations that are due to the finite number of
neurons.

Input correlations were substantially above the
autonomous case for correlated binary inputs, with
the magnitude of $\bar{C}$ decreasing
when the relative contribution of the recurrent activity
increased. This was the case for increasing
$R_{\rm a}$. The effect was opposite
for the Gaussian protocol, for which the input did
not induce correlations, but contributed to decorrelating
neural activity. In this case, the mean absolute correlation $\bar{C}$
was suppressed when the internal activity
became small in the limit $R_{\rm a}\to0$.
For larger $R_{\rm a}$, the recurrent input gained more
impact on neural activity relative to the external drive
and thus drove $\bar{C}$ towards an amount of
correlation that would be expected in the autonomous case.

\section{Discussion}
\label{sect:discussion}

The mechanisms for tuning the spectral radius via a
local homeostatic adaptation rule introduced in the present
study require neurons to have the ability to distinguish
and locally measure both external and recurrent input
contributions. For flow control, neurons need to be able
to compare the recurrent membrane potential
with their own activity, as assumed in
Section~\ref{sect_specrad_reg}. On the other hand,
variance control directly measures the variance of
the external input and derives the activity target variance
accordingly. The limiting factor to a successful spectral
radius control is the amount of cross-correlation 
induced by external driving statistics.
As such, the functionality and validity of the proposed
mechanisms depended on the ratio between external input,
i.e.\ feed-forward or feedback connections, with respect
to recurrent, or lateral connections. In general, it is not
straightforward to directly connect experimental evidence
regarding the ratio between recurrent and feed-forward
contributions to the effects observed in the model. It is,
however, worthwhile to note that the fraction of synapses
associated with interlaminar loops and intralaminar lateral
connections are estimated to make up roughly $50\%$
\citep{binzegger2004cortexcircuit}.
Relating this to our model, it implies that the significant
interneural correlations that we observed when external
input strengths were of the same order of magnitude as the
recurrent inputs, can not generally be considered an
artifact of biologically implausible parameter choices.
Synchronization \citep{echeveste2016drifting} is in
fact a widely observed phenomenon in the brain
\citep{Usrey_1999}, with possible relevance for 
information processing \citep{Salinas_2001}.

On the other hand, correlations due to shared input reduces the 
amount of information that can be stored in the neural ensemble
\citep{Bell_1995}. Maximal information is achieved if neural activities 
or spikes trains form an orthogonal ensemble 
\citep{Foeldiak_1990,Bell_1995,Tetzlaff2012}. Furthermore, neural 
firing in cortical microcircuits was found to be decorrelated across 
neurons, even if common external input was present \citep{Ecker2010},
that is, under a common orientation tuning. Therefore, the correlation we 
observed in our network due to shared input might be significantly reduced
by possible modifications/extensions of our model: First, a strict 
separation between inhibitory and excitatory nodes according to Dale's law
might help actively decorrelating neural activity \citep{Tetzlaff2012,Bernacchia2013}.
Second, if higher dimensional input was used, a combination of plasticity 
mechanisms in the recurrent and feed-forward connections could 
lead to the formation of an orthogonal representation of the input 
\citep{Foeldiak_1990,Bell_1995,Wick2010}, leading to richer, ``higher dimensional"
activity patterns, i.\ e.\ a less dominant largest principal component.
Ultimately, if these measures helped in reducing neural cross-correlations
in the model, we thus would expect them to also increase the accuracy of
the presented adaptation mechanisms. 
We leave these modifications to possible future research.

Overall, we found flow control to be generally more robust than
variance control in the sense that, while still being affected
by the amount of correlations within the neural reservoir,
the task performance was less so prone to changes in the
external input strength. Comparatively stable network
performance could be observed, in spite of certain
deviations from the desired spectral radius
(see Figure~\ref{fig:het_performance_sweep_flow_composite}).
A possible explanation may be that flow control uses
a distribution of samples from only a restricted part of
phase space, that is, from the phase space regions that
are actually visited  or ``used'' for a given input.
Therefore, while a spectral radius of unity ensures
--statistically speaking-- the desired scaling
properties in all phase-space directions, it seem to be
enough to control the correct scaling for the subspace
of activities that is actually used for a given set of
input patters. Variance control, on the other hand, relies 
more strictly on the assumptions that neural activities
are statistical independent. In consequence,
the desired results could only be achieved under a rather narrow
set of input statistics (independent Gaussian input with small
variance). In addition, the approximate expression derived for
the nonlinear transformation appearing in the mean field
approximation adds another layer of potential source of
systematic error to the control mechanism.
This aspect also speaks in favor of flow control, since its
rules are mathematically more simple. In contrast to variance
control, the stationarity condition stated in
Eq.~(\ref{flow_R_a_introduction}) is independent of the actual
nonlinear activation function used and could easily be adopted
in a modified neuron model. It should be noted, however, that
the actual target $R_{\rm t}$ giving optimal performance might
then also be affected.

Interestingly, flow control distinguishes
itself from a conventional local activity-target
perspective of synaptic homeostasis: There is no
predefined set point in Eq.~(\ref{a_i_flow_introduction}).
This allows heterogeneities of variances of neural activity
to develop across the network, while retaining the
average neural activity at a fixed predefined level.

We would like to point out that, for all the results presented here, 
only stationary processes were used for generating the input sequences. Therefore, it might
be worth considering the potential effects of non-stationary,
yet bounded, inputs on the results in future work. It should be noted,
however, that the temporal domain enters both adaptation
mechanisms only in the form of trailing averages of 
first and second moments. As a consequence, we expect the issue
of non-stationarity of external inputs to present itself
simply as a trade-off between slower adaptation,
i.e.\ longer averaging time scales, and the mitigation
of the effects of non-stationarities. Slow adaptation is,
however, completely in line with experimental results on
the dynamics of synaptic scaling, which is taking place
on the time scale of hours to days \citep{Turrigiano_1998,Turrigiano_2008}.

\section{Conclusion}
\label{sect:conclusion}

Apart from being relevant from a theoretical perspective,
we propose that the separability of recurrent and
external contributions to the membrane potential is an
aspect that is potentially relevant for the understanding
of local homeostasis in biological networks.
While homeostasis in neural compartments has been a
subject of experimental research
\citep{Chen_2008}, to our knowledge,
it has not yet been further investigated on a theoretical
basis, although it has been hypothesized that the functional 
segregation within the dendritic structure might also affect
(among other intraneural dynamical processes) homeostasis 
\citep{Narayanan2012}. 
The neural network model used in this
study lacks certain features characterizing biological
neural networks, like strict positivity of the neural firing
rate or Dale's law, viz E-I balance \citep{trapp2018ei}. 
Future research should therefore
investigate whether the here presented framework of
local flow control can be implemented within more
realistic biological neural network models. A particular concern
regarding our findings is that biological neurons are
spiking. The concept of an underlying instantaneous
firing rate is, strictly speaking, a theoretical
construct, let alone the definition of higher moments,
such as the ``variance of neural activity". It is
however acknowledged that the variability
of the neural activity is central for statistical
inference \citep{echeveste2020cortical}. It is also 
important to note that real-world biological control
mechanisms, e.g.\ of the activity, rely on physical
quantities that serve as measurable correlates. A well-known example
is the intracellular calcium concentration, which is essentially
a linearly filtered version of the neural spike train
\citep{Turrigiano_2008}. On a theoretical level,
Cannon and Miller showed that dual homeostasis
can successfully control the mean and variance of this
type of spike-averaging physical quantities
\citep{cannon2017stable}. An extension
of the flow control to filtered spike trains
of spiking neurons could be an interesting subject
of further investigations.
However, using spiking neuron models would have shifted
the focus of our research towards the theory of
liquid state machines \citep{Maass2002,Maass_2004},
exceeding the scope of this publication.
We therefore leave the extension to more
realistic network/neuron models to future work.

\section{Materials and methods}
\label{sect:methods}

\bigskip

\subsection{Model}
\label{sect:model}

We implemented an echo state network with $N$ neurons,
receiving $D_{\rm in}$ inputs. The neural activity is
$y_i\in[-1,1]$, $x_i$ the membrane potential, $u_i$ the
input activities, $W_{ij}$ the internal synaptic weights and $I_i$ the
external input received. The output layer will be specified later.
The dynamics
%
  \begin{equation}
  x_i(t) = a_i\sum_{j=1}^N W_{ij} y_j(t-1) + I_i(t),
  \qquad\quad
  y_i(t) = \tanh\left(x_i(t) - b_i\right)
  \label{x_i}
  \end{equation}
%
is discrete in time, where the input $I_i$ is treated
instantaneously. A tanh-sigmoidal has been used
as a nonlinear activation function.

The synaptic renormalization factor $a_i$ in (\ref{x_i})
can be thought of as a synaptic scaling parameter that
neurons use to regulate the overall strength of the
recurrent inputs. The strength of the inputs $I_i$ is
unaffected, which is biologically plausible
if external and recurrent signals arrive at separate
branches of the dendritic tree \citep{Spruston2008}.

The $W_{ij}$ are the bare synaptic weights, with $a_i
W_{ij}$ being the components of the effective weight matrix
$\widehat{W}_{\rm a}$. Key to our approach is that the
propagation of activity is determined by
$\widehat{W}_{\rm a}$, which implies that the spectral
radius of the effective, and not of the bare weight matrix needs
to be regulated.

The bare synaptic matrix $W_{ij}$ is sparse, with a
connection probability $p_{\rm r}=0.1$.
The non-zero elements are drawn from a Gaussian
with standard deviation
%
  \begin{equation}
  \sigma=\frac{\sigma_{\rm w}}{\sqrt{N p_{\rm r}}}\,,
  \label{sigma_w}
  \end{equation}
%
and vanishing mean $\mu$. Here $Np_{\rm r}$ corresponds
to the mean number of afferent internal synapses, with the
scaling $\sim 1/\sqrt{Np_{\rm r}}$ enforcing size-consistent
synaptic-weight variances. As discussed in the results section, 
we applied the following adaptation mechanisms:
%
  \begin{equation}
  b_i(t)= b_i(t-1) + \epsilon_{\rm b} \left[y_i(t) -
  \mu_{\rm t} \right] 
  \label{b_i}
  \end{equation}
%
for the thresholds $b_i$.
\begin{itemize}

\item Adaption of gains, using flow control:
%
  \begin{equation}
  a_i(t) = a_i(t-1)\Big[1+ \epsilon_{\rm a} \Delta
  R_i(t)\Big],
  \quad\quad
  \Delta R_i(t) = R_{\rm t}^2 {|y_i(t-1)|}^2 - {|x_{{\rm
        r},i}(t)|}^2\;.
  \label{a_i_flow}
  \end{equation}

%
\item Adaption of gains, with variance control:
  \begin{align}
  \label{a_i_variance_methods}
  a_i(t) &= a_i(t-1) + \epsilon_{\rm a} \left[
  \sigma_{{\rm t},i}^2(t) -
  {\left( y_i(t) - \mu^{\rm y}_i(t) \right)}^2\right] \\
  \label{sigm_target_methods}
  \sigma_{{\rm t},i}^2(t) &= 1 - \sqrt{1 + 2R_{\rm t}^2
    y_i(t)^2
    +
    2\sigma_{{\rm ext},i}^2(t)} \\
  \label{mu_y_methods}
  \mu^{\rm y}_i(t) &= \mu^{\rm y}_i(t-1) + \epsilon_\mu
  \left[y_i(t) - \mu^{\rm y}_i(t-1)\right] \\
  \label{sigm_ext_methods}
  \sigma_{{\rm ext},i}^2(t) &= \sigma_{{\rm ext},i}^2(t-1) +
  \epsilon_{\sigma} \left[\left(I_i(t) - \mu_{{\rm
      ext},i}(t)\right)^2 - \sigma_{{\rm ext},i}^2(t-1)\right] \\
  \label{mu_ext_methods}
  \mu_{{\rm ext},i}(t) &= \mu_{{\rm ext},i}(t-1) +
  \epsilon_\mu \left[I_i(t) - \mu_{{\rm ext},i}(t-1)\right] \; .
  \end{align}
%
Note that Eq.~(\ref{mu_y_methods})--(\ref{mu_ext_methods}) have the 
same mathematical form 
%
  \begin{equation*}
    \langle trail \rangle (t) = \langle trail \rangle (t-1) + 
    \epsilon\left[\langle var \rangle (t) - \langle trail \rangle (t-1)\right]
  \end{equation*}
%
since they only serve as trailing averages that are used in the two 
main equation (\ref{a_i_variance_methods}) and (\ref{sigm_target_methods}).
\end{itemize}
  
For a summary of all model parameters, see Table~\ref{tab_params}.

%
  \begin{table}[b]
    \centering
    \caption{Standard values for model parameters}
    \label{tab_params}
    \renewcommand{\arraystretch}{1.5}
    \begin{tabular}{ c|c|c|c|c|c|c|c }
      $N$ & $p_{\rm r}$ & $\sigma_{\rm w}$ & $\mu_{\rm
        t}$&$\epsilon_{\rm b}$
      & $\epsilon_{\rm a}$ & $\epsilon_\mu$ &
      $\epsilon_{\sigma}$ \\
      \hline
      $500$ & $0.1$ & $1$ & $0.05$ & $10^{-3}$ & $10^{-3}$ &
      $10^{-4}$ & $10^{-3}$
    \end{tabular}

  \end{table}
%

\bigskip

\subsection{Convergence acceleration 
	for flow control}
\label{sect:renorm_flow_control}

\begin{figure}
\begin{center}
\includegraphics[width=\textwidth]{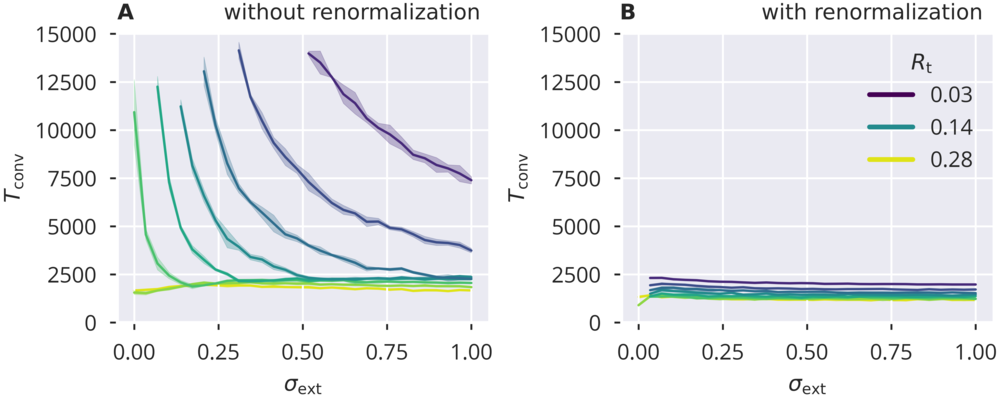}
\end{center}
\caption{{\bf Convergence time with and without adaptation rate
renormalization} Number of time steps $T_{\rm conv}$ needed for 
$|R_{\rm a}(t) - R_{\rm a}(t-1)|^2$ to fall below $10^{-3}$. 
Shown are results using heterogeneous Gaussian input without
and with, ({\bf A}) and respectively ({\bf B}), a
renormalization of the learning rate 
$\epsilon_a \rightarrow \epsilon_{\rm a} / \bar{x}^2_{\rm r}$. 
Note that, due to computational complexity, an estimate of
$R_{\rm a}$ given by (\ref{R_a}) was used. An initial offset 
of $0.5$ from the target $R_{\rm t}$ was used for all runs. 
Color coding of $R_{\rm t}$ is the same in both panels.}
\label{fig:flow_renorm}
\end{figure}

For small values of $R_{\rm t}$ and weak external input,
the average square activities and membrane potentials 
$y^2_i(t)$ and $x^2_{{\rm t},i}(t)$ can become very small. 
As a consequence, their difference entering 
$\Delta R_i(t)$ in (\ref{a_i_flow}) also becomes small 
in absolute value, slowing down the convergence process.
To eliminate this effect, we decided to rescale the learning 
rate by a trailing average of the squared recurrent 
membrane potential, i.\ e.\ 
$\epsilon_a \rightarrow \epsilon_{\rm a} / \bar{x}^2_{\rm r}$.
The effect of this renormalization is shown in 
Fig.~\ref{fig:flow_renorm}. Rescaling the learning rate effectively 
removes the significant rise of convergence times for small 
$\sigma_{\rm ext}$ and small $R_{\rm t}$.

\bigskip
\subsection{Input protocols}
\label{sect_input}

Overall, we examined four distinct input protocols.
%
  \begin{itemize}
    \item {\sl Homogeneous Gaussian.}
    Nodes receive inputs $I_i(t)$ that are drawn
    individually from a Gaussian with vanishing
    mean and standard deviation $\sigma_{\rm ext}$.

    \item {\sl Heterogeneous Gaussian.}
    Nodes receive stochastically independent inputs
    $I_i(t)$ that are drawn from Gaussian distributions
    with vanishing mean and node specific standard
    deviations $\sigma_{i, {\rm ext}}$.
    The individual $\sigma_{i, {\rm ext}}$ are normal
    distributed, as drawn from the positive part of
    a Gaussian with mean zero and variance
    $\sigma_{\rm ext}^2$.

    \item {\sl Homogeneous binary.}
    Sites receive identical inputs
    $I_i(t)=\sigma_{\rm ext} u(t)$,
    where $u(t)=\pm1$ is a binary input sequence.

    \item {\sl Heterogeneous binary.}
    We define with
    \begin{equation}
    I_i = W^{\rm u}_{i} u(t), \qquad\quad u_j(t)=\pm1
    \label{I_i}
    \end{equation}
    the afferent synaptic weight vector $W^{\rm u}_{i}$,
    which connects the binary input sequence $u(t)$ to the
    network. All $W^{\rm u}_{i}$ are drawn independently from a
    Gaussian with mean zero and standard deviation
    $\sigma_{\rm ext}$.
  \end{itemize}
%
The Gaussian input variant simulates external noise.
We used it in particular to test predictions of the
theory developed in Section~\ref{sect:MF_theory}.
In order to test the performance of the echo state
network
with respect to the delayed XOR task, the binary input
protocols are employed. A generalization of the here
defined protocols to the case of higher-dimensional input signals
would be straightforward.

\bigskip

\subsection{Spectral radius adaptation dynamics}
\label{sect_R_dynamics}

\begin{figure}[t]
    \includegraphics[width=1.0\textwidth]
    {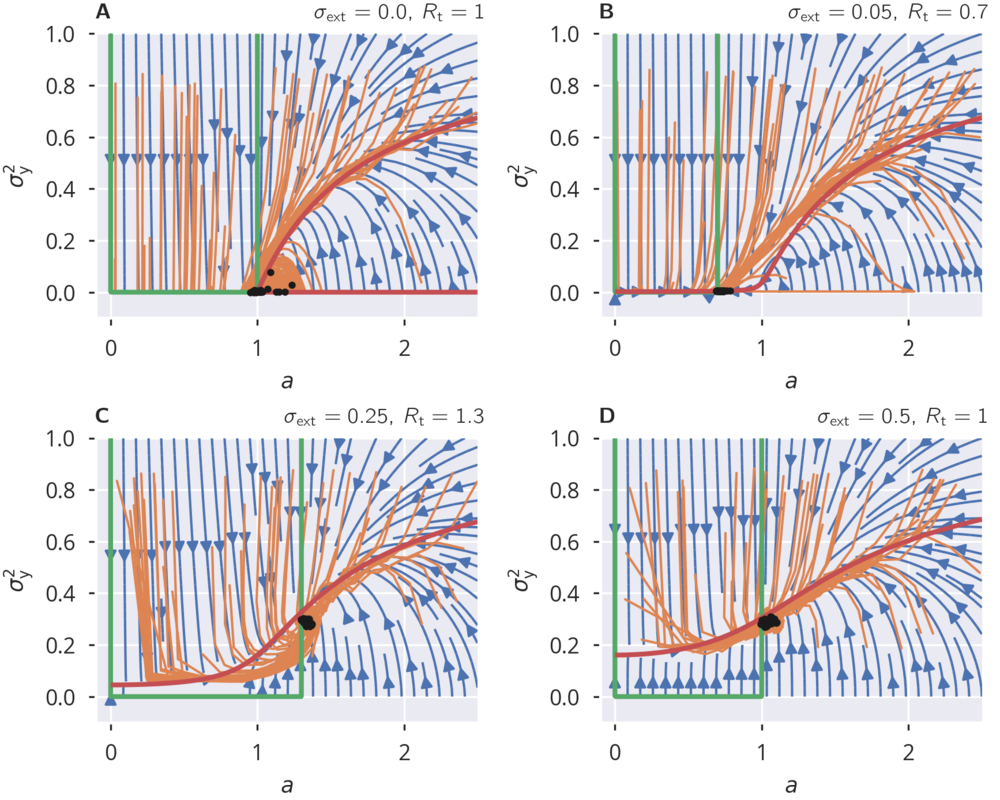}
    \caption{{\bf Spectral radius adaptation dynamics.}
      The dynamics of the synaptic rescaling factor $a$
      and the squared
      activity $\sigma_{\rm y}^2$ (orange), as given by
      (\ref{delta_R_global_introduction}), for $R_{\rm t}=1$.
      Also
      shown is
      the analytic approximation
      to the flow (blue), see (\ref{eq:gain_dyn_approx})
      and
      (\ref{eq:y_squ_dyn_approx}), and the respective
      nullclines
      $\Delta a=0$ (green) and $\Delta\sigma_{\rm y}^2=0$
      (red).
      For the input, the heterogeneous binary protocol is
      used. Panels {\bf A} to {\bf D} correspond to different
      combinations of external input strengths and target spectral radii.
      The black dots show the stead-state configurations of
      the simulated systems. $\epsilon_{\rm a} = 0.1$.}
    \label{fig_adaptation_dynamics}
  \end{figure}

For an understanding of the spectral radius adaptation
dynamics of flow control, it is of interest to examine the
effect of using the global adaptation constraint
%
  \begin{equation}
  \Delta R_i(t) = \frac{1}{N}\Big[
  R_{\rm t}^2\,{||\mathbf{y}(t-1)||}^2-
  {||\mathbf{x}_{\rm r}(t)||}^2 \Big]
  \label{delta_R_global}
  \end{equation}
%
in (\ref{a_i_flow_introduction}). The spectral radius condition
(\ref{flow_R_a_introduction}) is then enforced directly, with
the consequence that (\ref{delta_R_global}) is stable and precise
even in the presence of correlated neural activities (see
Figure~\ref{fig_R_a_regulation}C). This rule, while not
biologically plausible, provides an opportunity to examine
the dynamical flow, besides the resulting state.
There are two dynamic variables, $a = a_i \; \forall i$,
where, for the sake of simplicity, we assumed that all $a_i$ 
are homogeneous,
and the activity variance $\sigma_{\rm y}^2=||\mathbf{y}||^2/N$.
The evolution of $(a,\sigma_{\rm y}^2)$  resulting from
the global rule (\ref{delta_R_global_introduction}) is shown in
Figure~\ref{fig_adaptation_dynamics}.
%
%
For the flow, $\Delta a= a(t+1)-a(t)$ and
$\Delta\sigma_{\rm y}^2 = \sigma_{\rm y}^2(t)
- \sigma_{\rm y}^2(t-1)$,
the approximation
%
  \begin{align}
  \label{eq:gain_dyn_approx}
  \Delta a &= \epsilon_{\rm a} a
  \left(R_{\rm t}^2-a^2\sigma^2_{\rm w}\right) \sigma_{\rm y}^2
  \\
  \Delta\sigma_{\rm y}^2 &=
  1 - \sigma_{\rm y}^2
  - \frac{1}{\sqrt{1+2a^2 \sigma^2_{\rm w} \sigma_{\rm y}^2 + 2 \sigma_{\rm
        ext}}}
  \label{eq:y_squ_dyn_approx}
  \end{align}
%
is obtained. For the scaling factor $a$, this leads to a
fixed point of $R_{\rm t}/\sigma_{\rm w}$. We used
the mean-field approximation for neural variances that is
derived in Section~\ref{sect:MF_theory}.
The analytic flow compares well with numerics, as shown in
Figure~\ref{fig_adaptation_dynamics}. For a subcritical rescaling
factor $a$ and $\sigma_{\rm ext}=0$,
the system flows towards a line of fixpoints defined by a
vanishing $\sigma_{\rm y}^2$ and a finite $a\in[0,1]$,
see Figure~\ref{fig_adaptation_dynamics}A.
When starting with $a>0$, the fixpoint is instead
$(a,\sigma_{\rm y}^2)=(1,0)$. The situation changes
qualitatively for finite external inputs, viz when
$\sigma_{\rm ext}>0$, as shown in Figure~\ref{fig_adaptation_dynamics}B--D.
The nullcline $\Delta\sigma_{\rm y}^2=0$ is now
continuous and the system flows to the fixed point, as shown in
Figure~\ref{fig_adaptation_dynamics}B--D,
with the value of $\sigma_{\rm y}^2$ being
determined by the intersection of the two nullclines. In addition,
we also varied the target spectral radius,
see Figure~\ref{fig_adaptation_dynamics}B/C. This caused a slight
mismatch between the flow of the simulated systems and the analytic flow.
It should be noted, however, that this is to be expected anyhow
since we used an approximation for the neural variances,
again, see Section~\ref{sect:MF_theory}.

This analysis shows that external input is necessary for a robust
flow towards the desired spectral weight, the reason being
that the dynamics dies out before the spectral weight
can be adapted when the isolated systems starts in the
subcritical regime.

\bigskip

\subsection{Extended theory of flow control for independent neural activity}
\label{sect:flow_theo}

We would like to show that the stationarity condition
in Eq.~(\ref{flow_R_a_introduction}) results in the correct
spectral radius, under the special case of independently
identically distributed neural activities with zero mean.

We start with Eq.~(\ref{flow_R_a_introduction}) as a
stationarity condition for a given $R_{\rm t}$:
%
  \begin{equation}
  {\big\langle {||\mathbf{x}_{\rm r}(t)||}^2 \big\rangle}_{\rm
    t}
  \overset{!}{=}
  R_{\rm t}^2 {\big\langle
    \,{||\mathbf{y}(t-1)||}^2\big\rangle}_{\rm t} \, .
\label{eq:stat_cond}
  \end{equation}
%
We can express the left side of the equation as
%
  \begin{equation}
  \mathrm{E}\left[ \mathbf{y}^\dagger (t)
  \widehat{W}_{\rm a}^\dagger \widehat{W}_{\rm a}
  \mathbf{y}(t) \right]_t \, .
  \end{equation}
%
We define $\widehat{U}_{\rm a}
\equiv= \widehat{W}_{\rm a}^\dagger \widehat{W}_{\rm a}$
with $\{\sigma^2_k\}$ being the set of eigenvalues,
which are also the squared singular values of
$\widehat{W}_{\rm a}$, and $\{\mathbf{u}_k\}$ the
respective set of orthonormal (column) eigenvectors.
We insert the identity
$\sum_{k=1}^N \mathbf{u}_k \mathbf{u}^\dagger_k$ and find
%
\begin{align}
& \mathrm{E}\left[ \mathbf{y}^\dagger (t)
\widehat{U}_{\rm a}
\sum_{k=1}^N \mathbf{u}_k \mathbf{u}^\dagger_k
\mathbf{y}(t) \right]_t \\
= & \mathrm{E}\left[ \sum_{k=1}^N \sigma^2_k
\mathbf{y}^\dagger (t) \mathbf{u}_k
\mathbf{u}^\dagger_k \mathbf{y}(t) \right]_t \\
= & \sum_{k=1}^N \sigma^2_k
\mathbf{u}^\dagger_k
\mathrm{E}\left[\mathbf{y}(t)\mathbf{y}^\dagger
(t)\right]_t
\mathbf{u}_k \\
= & \sum_{k=1}^N \sigma^2_k
\mathbf{u}^\dagger_k
\widehat{C}_{\rm yy}
\mathbf{u}_k \\
= & \mathrm{Tr}\left(
\widehat{D}_{\sigma^2} \widehat{S}^\dagger_{\rm u}
\widehat{C}_{\rm yy} \widehat{S}_{\rm u} \right) \, .
\end{align}
%
Given zero mean neural activity, $\widehat{C}_{\rm yy} =
\mathrm{E}[\mathbf{y}(t)\mathbf{y}^\dagger (t)]_t$
is the covariance matrix of neural activities.
$\widehat{D}_{\sigma^2}$ is a diagonal matrix holding
the $\{\sigma^2_k\}$ and $\widehat{S}_{\rm u}$ is a
unitary matrix whose columns are $\{\mathbf{u}_k\}$.
$\widehat{S}^\dagger_{\rm u}
\widehat{C}_{\rm yy} \widehat{S}_{\rm u}$ is expressing
$\widehat{C}_{\rm yy}$ in the diagonal basis of
$\widehat{U}_{\rm a}$.

Including the right hand side of (\ref{eq:stat_cond}), we get
%
  \begin{equation}
  \mathrm{Tr}\left(
  \widehat{D}_{\sigma^2} \widehat{S}^\dagger_{\rm u}
  \widehat{C}_{\rm yy} \widehat{S}_{\rm u} \right)
  = R^2_{\rm t} \mathrm{Tr}\left(\widehat{C}_{\rm
    yy}\right) \, .
  \end{equation}
%
However, since the trace is invariant under a change of
basis, we find
%
  \begin{equation}
  \mathrm{Tr}\left(
  \widehat{D}_{\sigma^2} \widehat{S}^\dagger_{\rm u}
  \widehat{C}_{\rm yy} \widehat{S}_{\rm u} \right)
  = R^2_{\rm t} \mathrm{Tr}\left(
  \widehat{S}^\dagger_{\rm u} \widehat{C}_{\rm yy}
  \widehat{S}_{\rm u} \right) \, .
  \end{equation}
%
Defining $\widehat{C}^{\rm u} \equiv=
\widehat{S}^\dagger_{\rm u} \widehat{C}_{\rm yy}
\widehat{S}_{\rm u}$,
we get
%
  \begin{equation}
  \sum_{k=1}^N \sigma^2_k C^{\rm u}_{kk}
  = R^2_{\rm t} \sum_{k=1}^N C^{\rm u}_{kk} \, .
  \label{eq:sum_sv_base_transf}
  \end{equation}
%
If we assume that the node activities are independently
identically distributed with zero mean, we get
$(\widehat{C}_{\rm yy})_{ij} =
(\widehat{C}^{\rm u} )_{ij} =
\left\langle y^2\right\rangle_{t} \delta_{ij}$.
In this case, which was also laid out in Section~\ref{sec:sing_values},
the equation reduces to
%
  \begin{equation}
  \sum_{k=1}^N \sigma^2_k  = R^2_{\rm t} N  \, .
  \label{eq:sum_sv_specrad}
  \end{equation}
%
The Frobenius norm of a square Matrix $\widehat{A}$ is
given by
${\lVert \widehat{A} \rVert}^2_{\rm F}
\equiv = \sum_{i,j} \widehat{A}^2_{ij}$. Furthermore,
the Frobenius norm is linked to the singular values via
${\lVert \widehat{A} \rVert}^2_{\rm F} =
\sum_k \sigma^2_k (\widehat{A})$
\citep{sengupta1999distributions,shen2001singular}.
This allows us to state
%
  \begin{equation}
  \sum_{i,j} {\left(\widehat{W}_{\rm a}\right)}^2_{ij}  =
  R^2_{\rm t} N
  \end{equation}
%
which, by using (\ref{R_a}), gives
%
  \begin{equation}
  R^2_{\rm a}  = R^2_{\rm t}  \, .
  \end{equation}
%
A slightly less restrictive case is that of uncorrelated
but inhomogeneous activity, that is
${(\widehat{C}_{\rm yy})}_{ij} =
{\left\langle y_i^2 \right\rangle}_{t} \delta_{ij}$.
The statistical properties of the diagonal
elements $C^{\rm u}_{kk}$ then determine to which degree
one can still claim that Eq.~(\ref{eq:sum_sv_base_transf})
leads to Eq.~(\ref{eq:sum_sv_specrad}).
Figure~\ref{S7_Fig} in the supplementary materials
shows an example of a randomly generated realization of
${(\widehat{C}_{\rm yy})}_{ij} =
{\left\langle y_i^2\right\rangle}_{t}$
and the resulting diagonal elements of $\widehat{C}^{\rm u}$,
where the corresponding orthonormal basis $\widehat{S}_{\rm u}$
was generated from the SVD of a random Gaussian matrix.
As one can see,
the basis transformation has a strong smoothing effect
on the diagonal entries, while the mean over the diagonal
elements is preserved. Note that this effect was not
disturbed by introducing random row-wise multiplications to the
random matrix from which the orthonormal basis was
derived. The smoothing of the diagonal entries allows us to
state that
$C^{\rm u}_{kk} \approxeq \left\langle y^2\right\rangle$
is a very good approximation in the case considered,
which therefore reduces (\ref{eq:sum_sv_base_transf})
to the homogeneous case previously described.
We can conclude that the adaptation mechanism
also gives the desired spectral radius under uncorrelated
inhomogeneous activity.

In the most general case, we can still state that if
$C^{\rm u}_{kk}$ and $\sigma^2_k$ are uncorrelated,
for large $N$, Eq.~(\ref{eq:sum_sv_base_transf}) will
tend towards
%
  \begin{equation}
  N \left\langle \sigma^2\right\rangle
  \left\langle C^{\rm u}\right\rangle
  = N R^2_{\rm t} \left\langle C^{\rm u}\right\rangle
  \end{equation}
%
which would also lead to Eq.~(\ref{eq:sum_sv_specrad}).
However, we can not generally guarantee statistical
independence since the recurrent contribution on
neural activities and the resulting
entries of $\widehat{C}_{\rm yy}$ and thus also
$C^{\rm u}_{kk}$ are linked to $\widehat{S}$
and $\sigma^2_k$, being the SVD of the recurrent
weight matrix.

\bigskip
\bigskip

\subsection{Mean field theory for echo state layers}
\label{sect:MF_theory}

In the following, we deduce analytic expressions allowing
to examine the state of echo-state layers subject to a
continuous timeline of inputs. Our approach is similar to
the one presented by \citet{Massar2013}.

The recurrent part of the input $x_i$ received by a
neuron is a superposition of $Np_{\rm r}$ terms, which are
assumed here to be uncorrelated. Given this assumption,
the self-consistency equations
%
  \begin{align}
  \label{self_consistency_sigma_t}
  \sigma_{{\rm y},i}^2&=\int_{-\infty}^{\infty}
  {\rm dx}\tanh^2(x) N_{\mu_i,\sigma_i}(x) - \mu^2_{{\rm y},i}
  \\
  \label{self_consistency_mu_t}
  \mu_{{\rm y},i} &= \int_{-\infty}^{\infty}
  {\rm dx}\tanh(x) N_{\mu_i,\sigma_i}(x)
  \\
  \sigma^2_i&=a_i^2\sigma_{\rm w}^2\left\langle\sigma_{{\rm
      y},j}^2\right\rangle_j+
  \sigma_{{\rm ext},i}^2, \qquad\quad
  \mu_i = \mu_{{\rm ext},i} - b_i
  \label{self_consistency_sigma_mu}
  \end{align}
%
determine the properties of the stationary state.
We recall that $\sigma_{\rm w}$
parameterizes the distribution of bare synaptic
weights via (\ref{sigma_w}). The general expressions
(\ref{self_consistency_sigma_t}) and
(\ref{self_consistency_mu_t}) hold for all neurons,
with the site-dependency entering
exclusively via $a_i$, $b_i$, $\sigma_{{\rm ext},i}$
and $\mu_{{\rm ext},i}$, as in
(\ref{self_consistency_sigma_mu}),
with the latter characterizing the standard deviation and
the mean of the input. Here,
$a_i^2\sigma_{\rm w}^2\sigma_{\rm y}^2$ is the variance
of the recurrent contribution to the membrane potential,
$x$, and $\sigma^2$ the respective total variance.
The membrane potential is Gaussian distributed, as
$N_{\mu,\sigma}(x)$, with mean $\mu$ and variance
$\sigma^2$, which are both to be determined
self-consistently.
Variances are additive for stochastically independent
processes, which has been assumed in
(\ref{self_consistency_sigma_mu}) to be the case for
recurrent activities and the external inputs.
The average value for the mean neural activity
is $\mu_i$.

For a given set of $a_i$, $\sigma_{{\rm ext},i}$ and $b_i$, the
means and variances of neural activities, $\sigma^2_{{\rm y},i}$
and $\mu_{{\rm y},i}$, follow implicitly.

\begin{figure}[t]
    \includegraphics[width=1.0\textwidth]
    {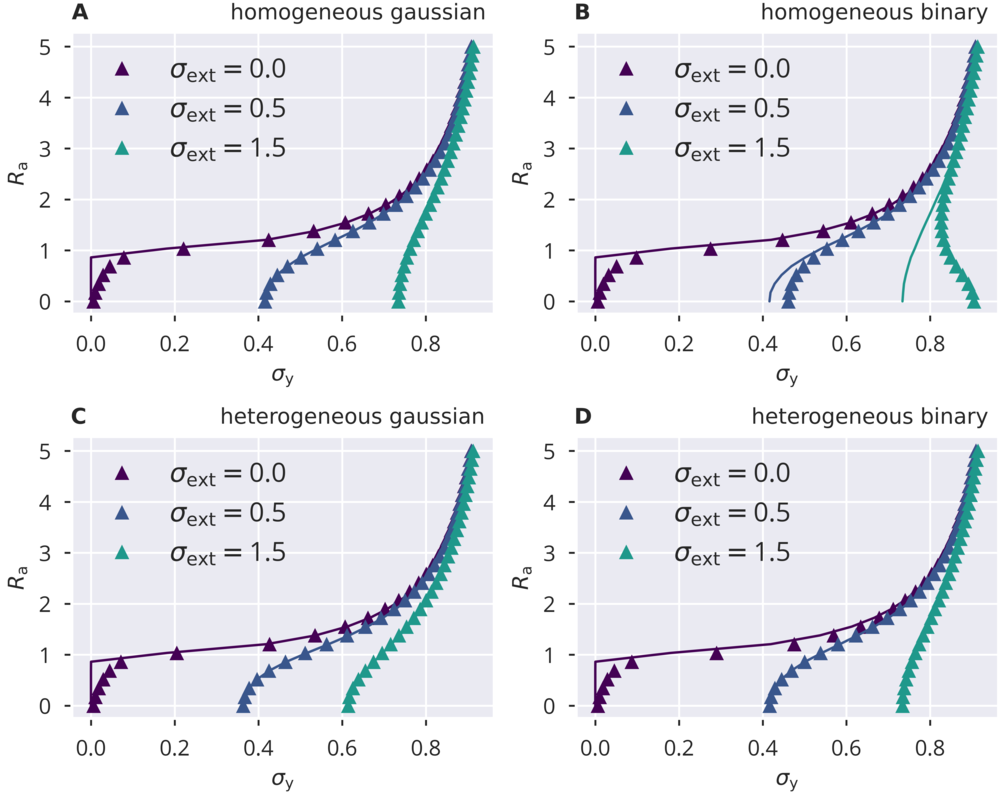}
    \caption{{\bf Variance control for the spectral
        radius.}
      The spectral radius $R_{\rm a}$, given by the
      approximation
      $R_{\rm a}^2=\sum_ia_i^2/N$, for the four input
      protocols
      defined in Section~\ref{sect_input}.
      Lines show the numerical self-consistency solution
      of
      (\ref{self_consistency_sigma_t}), symbols the full
      network
      simulations.
      Note the instability for small $\sigma_{\rm y}$ and
      $\sigma_{\rm ext}$.
      {\bf A}: Homogeneous independent Gaussian input.
      {\bf B}: Homogeneous identical binary input.
      {\bf C}: Heterogeneous independent Gaussian input.
      {\bf D}: Heterogeneous identical binary input.
    }
    \label{fig_R_a_input_protocols}
  \end{figure}

We compared the numerically determined solutions of
(\ref{self_consistency_sigma_t}) and
(\ref{self_consistency_mu_t}) against full
network simulations using, as throughout this study,
$N=500$, $p_{\rm r}=0.1$,
$\sigma_{\rm w}=1$, $\mu_{\rm t}=0.05$.
In Figure~\ref{fig_R_a_input_protocols}, the spectral
radius $R_{\rm a}$ is given for the four input protocols defined
in Section~\ref{sect_input}.
The identical ensemble of input standard
deviations $\sigma_{{\rm ext},i}$ enters both theory
and simulations.
%
%
Theory and simulations are in good accordance for
vanishing input. Here, the reason is that finite
activity levels are sustained in an autonomous
random neural network when the ongoing dynamics
is chaotic and hence decorrelated. For reduced
activity levels, viz for small variances
$\sigma_{\rm y}^2$, the convergence of the network dynamics
is comparatively slow, which leads to a certain
discrepancy with the analytic prediction
(see Figure~\ref{fig_R_a_input_protocols}).

\bigskip

\subsubsection{Gaussian~approximation}
\label{sec:Gaussian_Approximation}

The integral occurring in the self-consistency condition
(\ref{self_consistency_sigma_t}) can be evaluated
explicitly when a tractable approximation to the squared transfer
function $\tanh^2()$ is available. A polynomial approximation
would capture the leading behavior close to the origin, however
without accounting for the fact that $\tanh^2()$
converges to unity for large absolute values of the membrane
potential. Alternatively, an approximation incorporating both
conditions, the correct second-order scaling for small, and the
correct convergence for large arguments, is given by the
Gaussian approximation
%
  \begin{equation}
  \tanh^2(x) \approx 1 - \exp\left(-x^2 \right)\,.
  \label{gaussian_approximation}
  \end{equation}
%
With this approximation the integral in
(\ref{self_consistency_sigma_t}) can be evaluated
explicitly. The result is
%
  \begin{align}
  \frac{1}{1-\sigma^2_{\rm y} - \mu^2_{\rm y}}
  &= \sqrt{1+2\sigma^2}
  / \exp\left(-\mu^2/\left(1 + 2 \sigma^2\right) \right)
  \label{selfConsistency_GaussianApprox} \\
  &= \sqrt{1+2a^2\sigma^2_{\rm w}
    \sigma^2_{\rm y} + 2\sigma^2_{\rm ext} }
  / \exp\left(-\mu^2/\left(1 + 2a^2\sigma^2_{\rm w}
  \sigma^2_{\rm y} + 2\sigma^2_{\rm ext}\right) \right)
  \,.
  \nonumber
  \end{align}
%
Assuming that $\mu \approx 0$ and $\mu_{\rm y} \approx
0$, inverting the first equation yields a relatively simple analytic
approximation for the variance self-consistency equation:
%
  \begin{equation}
  \sigma^2_{\rm y} = 1 - \frac{1}{\sqrt{1+2a^2\sigma^2_{\rm w}
      \sigma^2_{\rm y} + 2\sigma^2_{\rm ext} }} \; .
  \label{eq:sigm_y_approx}
  \end{equation}
%
This equation was then used for the approximate update rule in
(\ref{sigm_target}) and (\ref{eq:y_squ_dyn_approx}).

\begin{figure}[t]
\centering
\includegraphics[width=\textwidth]{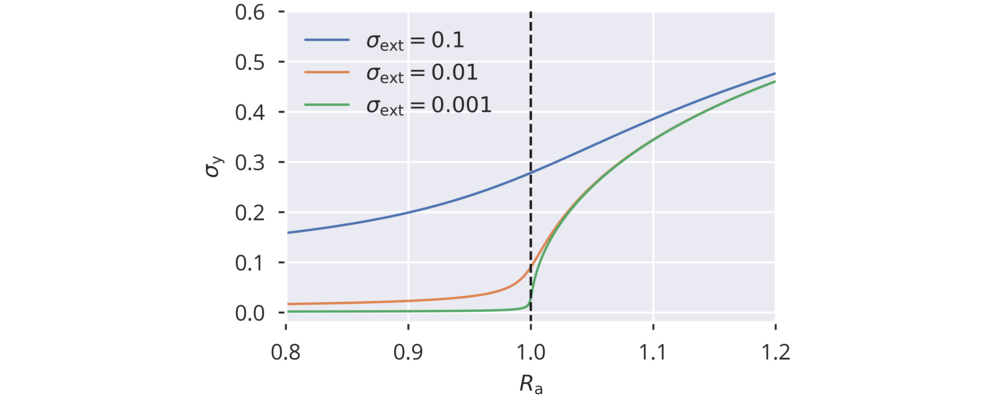}
\caption{{\bf Phase transition of activity variance} Shown are solutions of
the analytical approximation given in (\ref{eq:sigm_y_approx_self_consist}),
capturing the onset of activity (characterized by its variance 
$\sigma^2_{{\rm y}}$) at the critical point $R_{\rm a}=1$.}
\label{fig:phase_trans_analytic}
\end{figure}

Alternatively, we can write (\ref{eq:sigm_y_approx}) as a self-
consistency equation between $\sigma_{{\rm y}}^2$, $\sigma_{{\rm ext}}^2$
$a^2\sigma_{{\rm w}}^2 = R^2_{\rm a}$, describing a phase transition at
$R_{\rm a} = 1$:
%
\begin{equation}
 2R^2_{\rm a} \sigma_{{\rm y}}^2 \left(1 - \sigma_{{\rm y}}^2 \right)^2
= 1 - \left(1+2 \sigma_{{\rm ext}}^2\right)\left(1- \sigma_{{\rm y}}^2\right)^2 \; .
\label{eq:sigm_y_approx_self_consist}
\end{equation}
%
See Fig.~\ref{fig:phase_trans_analytic} for solutions of 
(\ref{eq:sigm_y_approx_self_consist}) for different values of
$\sigma^2_{{\rm ext}}$. Note that for vanishing
external driving and values of $R_{\rm a}$ above but close to 
the critical point, the standard deviation
$\sigma_{{\rm y}}$ scales with $\sigma_{{\rm y}} \propto (R_{\rm a} - 1)^{1/2}$,
which is the typical critical exponent for the order parameter in classical
Landau theory of second-order phase transitions \citep[p. 169]{Gros_ComplexSystems}.
If combined with a slow homeostatic process, flow or variance
control in our case, this constitutes a system with an absorbing
phase transition \citep[p. 182-183]{Gros_ComplexSystems}, settling
at the critical point $R_{\rm a}  = 1$.
This phase transition can
also be observed in Fig.~\ref{fig_R_a_input_protocols} for 
$\sigma_{{\rm ext}} = 0$ as a sharp onset in $\sigma_{{\rm y}}$.
%
%

\section*{Conflict of Interest Statement}

The authors declare that the research was conducted in
the absence of any commercial or financial relationships
that could be construed as a potential conflict of interest.

\section*{Author Contributions}

Both authors, F.S. and C.G., contributed equally to the
writing and review of the manuscript. F.S. provided the code,
ran the simulations and prepared the figures.


\section*{Acknowledgments}

The authors acknowledge the financial support of
the German research foundation (DFG) and discussions
with R.~Echeveste. This manuscript was published 
as a pre-print on biorxiv \citep{Schubert2020_biorxiv}.

\section*{Data Availability Statement}

The datasets generated for this study can be found in 
\url{https://itp.uni-frankfurt.de/~fschubert/data_esn_frontiers/}.

Simulation and plotting code is available in 
\url{https://github.com/FabianSchubert/ESN_Frontiers}.

\newpage
\section{Supplementary Figures}




\begin{figure}[htbp]
  \begin{center}
    \includegraphics[width=\textwidth]{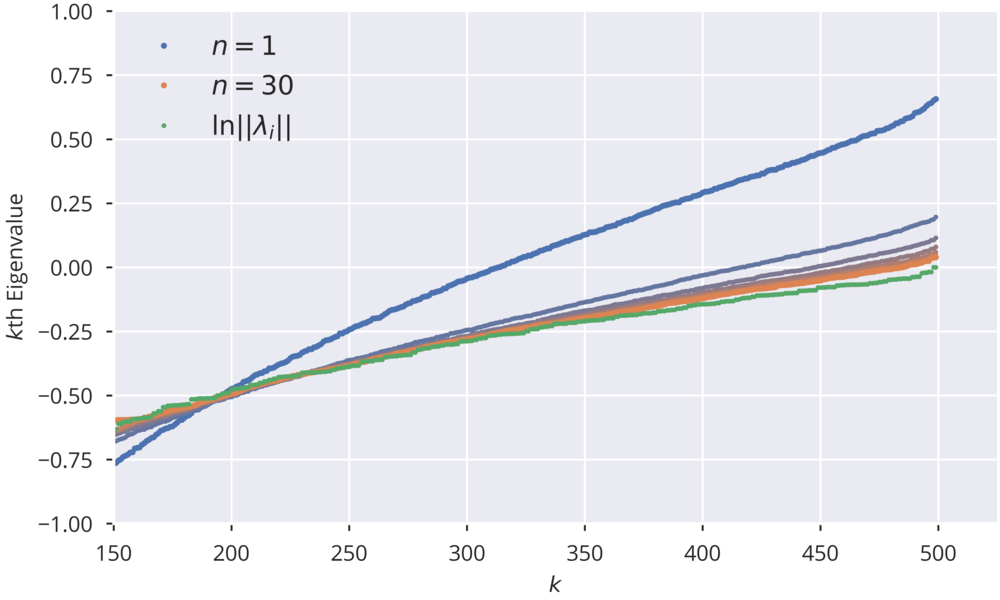}
  \end{center}
  \caption{{\bf Convergence of Lyapunov Spectrum.}
    Convergence of eigenvalues of 
    $\ln\big((\widehat{W}^n)^\dagger\widehat{W}^n 
    \big)/(2n)$ 
    for different $n$, as discussed in 
    Section~\ref{sec:sing_values}.
    $\widehat{W}$ is a random Gaussian matrix which 
    was rescaled to a spectral radius of one. 
    Colors from blue to orange encode the exponent $n$
    ranging between $1$ and $30$. Green dots 
    show the theoretical limit of 
    $\ln\lVert \lambda_i \rVert$, where $\lambda_i$
    is the $i$th eigenvalue of $\widehat{W}$.}
  \label{S3_Fig}
\end{figure}

\begin{figure}[htbp]
  \begin{center}
    \includegraphics[width=1.0\textwidth]
    {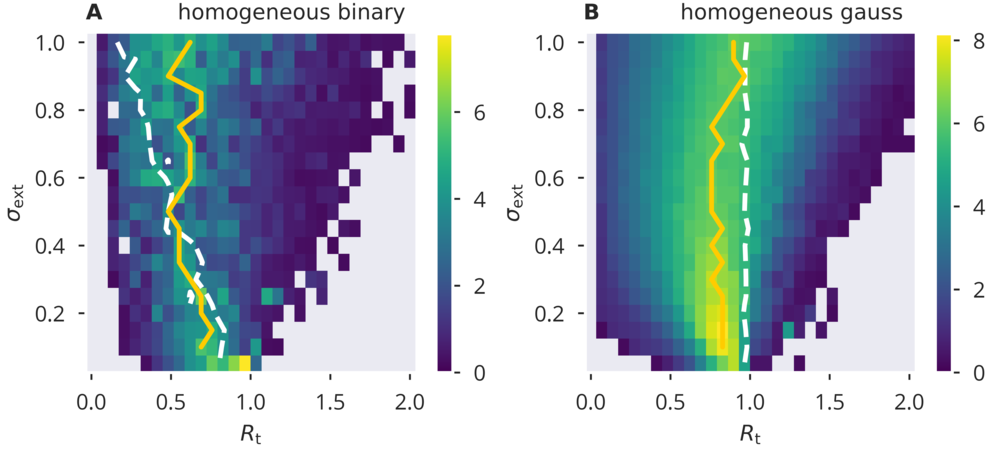}
  \end{center}
  \caption{{\bf XOR performance for flow control, 
    homogeneous input.} 
    Numerical results for the network performance under a 
    time-delayed XOR task, as defined in 
    Section~\ref{sect:XOR}, 
    using homogeneous binary/Gaussian input.  
    Shown are color-coded performance sweeps for the 
    XOR-performance 
    (\ref{MC_XOR}), averaged over five trials. The input
    has variance $\sigma_{\rm ext}^2$ and
    the target for the spectral radius $R_{\rm t}$.
    A/B panels are for binary/Gaussian 
    input protocols. Optimal performance for a given $\sigma_{\rm 
      ext}$ is given by yellow solid lines, measured value of 
    $R_{\rm a} = 1$ by white dashed lines.}
  \label{S4_Fig}
\end{figure}

\begin{figure}[htbp]
  \begin{center}
    \includegraphics[width=1.0\textwidth]
    {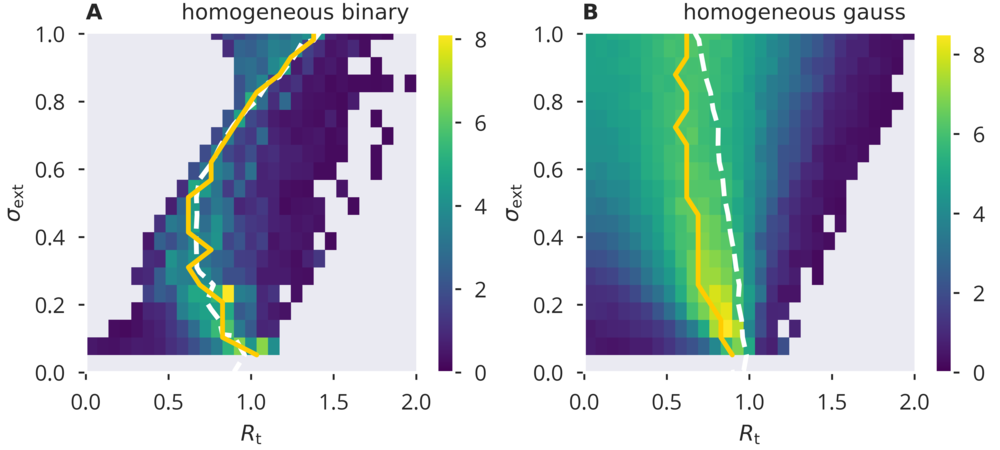}
  \end{center}
  \caption{{\bf XOR performance for variance control, 
    homogeneous input.} 
    Numerical results for the network performance under a 
    time-delayed XOR task, as defined in 
    Section~\ref{sect:XOR}, 
    using homogeneous binary/Gaussian input.  
    Shown are color-coded performance sweeps for the 
    XOR-performance 
    (\ref{MC_XOR}), averaged over five trials. The input
    has variance $\sigma_{\rm ext}^2$ and
    the target for the spectral radius $R_{\rm t}$.
    A/B panels are for binary/Gaussian 
    input protocols. Optimal performance for a given $\sigma_{\rm 
    ext}$ is given by yellow solid lines, measured value of $R_{\rm 
    a} = 1$ by white dashed lines.}
  \label{S5_Fig}
\end{figure}

\begin{figure}[htbp]
  \begin{center}
    \includegraphics[width=1.0\textwidth]
    {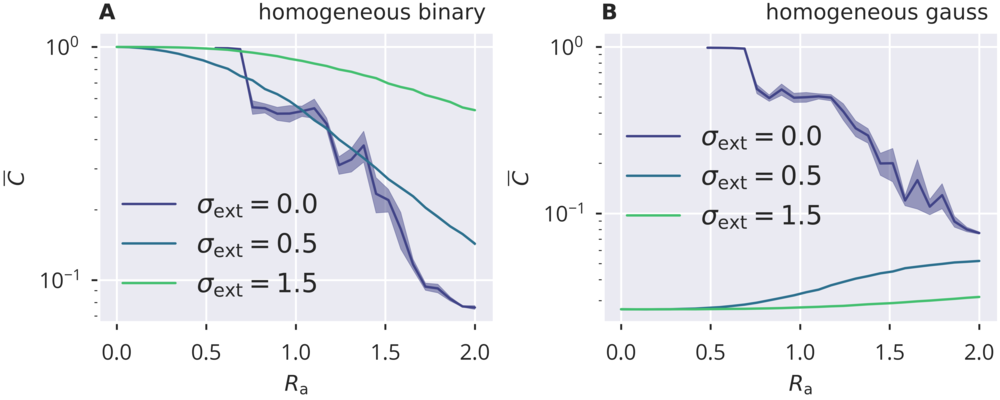}
  \end{center}
  \caption{{\bf Input induced activity correlations.} 
    For homogeneous binary and Gaussian inputs
    (A/B), the dependency of mean 
    activity 
    cross correlations $\bar{C}$, see 
    Eq.~(\ref{crossCorr}).
    $\bar{C}$ is 
    shown as a function of the target spectral radius 
    $R_{\rm a}$. Results are obtained
    for $N\!=\!500$ sites by averaging over five trials,
    with shadows indicating the accuracy. Correlations 
    are 
    due to finite-size effect for the autonomous case 
    $\sigma_{\rm ext}\!=\!0$.}
  \label{S6_Fig}
\end{figure}

\begin{figure}[htbp]
  \begin{center}
    \includegraphics[width=1.0\textwidth]{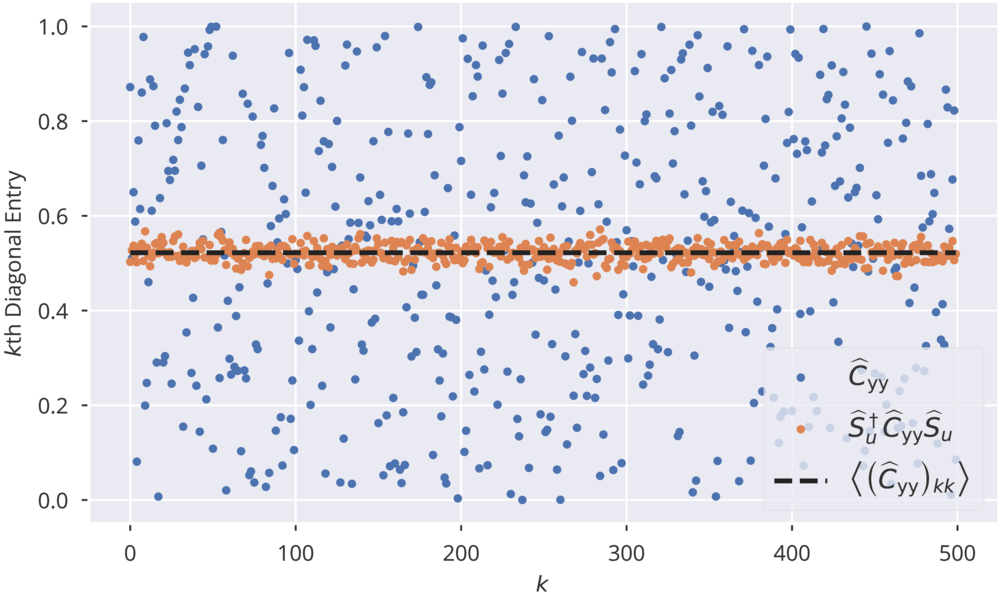}
  \end{center}
  \caption{{\bf Diagonal Elements of a randomly 
    generated 
    covariance matrix and its representation in the 
    $\mathbf{\widehat{S}_{\rm u}}$ basis.} 
    $\widehat{C}_{\rm yy}$ is a diagonal matrix with 
    diagonal entries randomly drawn from $[0,1]$, 
    $\widehat{S}_{\rm u}$ is the orthonormal eigenbasis 
    of $\widehat{W}_{\rm a}^\dagger \widehat{W}_{\rm 
    a}$, 
    where $\widehat{W}_{\rm a}$ is a random Gaussian 
    matrix. 
    The black dashed line denotes the average over the 
    diagonal entries of $\widehat{C}_{\rm yy}$.}
  \label{S7_Fig}
\end{figure}

\end{document}